\documentclass[12pt]{iopart}

\usepackage{graphicx}
\begin{document}

\def\ket#1{|#1\rangle}
\def\bra#1{\rangle#1|}

\def\lsim{\;\raisebox{-.6ex}{$\stackrel{<}{\sim}$}\;}

\title{Supernova Signatures of Neutrino Mass Ordering }

\author{Kate Scholberg}

\address{Duke University Physics Department, Durham, NC 27708, United States of America}
\ead{schol@phy.duke.edu}
\vspace{10pt}
\begin{indented}
\item[]December 2017
\end{indented}

\begin{abstract}
A suite of detectors around the world is poised to measure the flavor-energy-time evolution of the ten-second burst of neutrinos from a core-collapse supernova occurring in the Milky Way or nearby.  Next-generation detectors to be built in the next decade will have enhanced flavor sensitivity and statistics.   Not only will the observation of this burst allow us to peer inside the dense matter of the extreme event and learn about the collapse processes and the birth of the remnant, but the neutrinos will bring information about neutrino properties themselves.  This review surveys some of the physical signatures that the currently-unknown neutrino mass pattern will imprint on the observed neutrino events at Earth, emphasizing the most robust and least model-dependent signatures of mass ordering.
\end{abstract}

%
%
%
%
%

\section{Introduction}

The observation of the burst of neutrinos from Supernova 1987A~\cite{Bionta:1987qt,Hirata:1987hu,Alekseev:1987ej} in the Large Magellanic Cloud just outside our Milky Way galaxy confirmed the basic picture of core-collapse supernovae, but also brought new knowledge about neutrinos themselves.  
The 1987A neutrino signal in water and scintillator detectors led to the best limits, at that time, on absolute mass scale of the neutrino, based on the lack of energy-dependent spread (e.g., \cite{Schramm:1990pf}).  These limits were soon exceeded by terrestrial measurements, but other limits on neutrino properties (and other particle physics) still stand as the most stringent~\cite{Schramm:1990pf, Raffelt:1998hw}.  Statistics for the 1987A observation were paltry though--- just a few dozen events were recorded, nearly all likely to be electron antineutrinos~\cite{Vissani:2014doa}.

A new generation of neutrino detectors stands ready for the next burst, and a future generation of detectors is under design and construction~\cite{Scholberg:2012id}.
The next observed core-collapse burst, with much higher statistics and greater flavor sensitivity, will lead to a spurt of progress in understanding of core-collapse mechanisms and remnants.  In addition, as for SN1987A,  it will also lead to new knowledge about the nature of neutrinos.

Since SN1987A we have learned a tremendous amount about neutrinos.  Many experiments using a variety of neutrino sources have told us that neutrinos have mass and oscillate, and a three-mass-state/three-flavor-state picture fits nearly all of the data very well~\cite{Gonzalez-Garcia:2015qrr}.  There are still unknowns, however, and a supernova neutrino burst may tell us about some of these unknowns.  While laboratory measurements will likely address many of these unknowns in due course,  a timely supernova burst may be the first to give us some of the answers.  Even if terrestrial measurements come first, they will help to constrain the observables to improve astrophysical interpretation of the data.  Better astrophysical observations of the supernova (in electromagnetic wavelengths and potentially in gravitational waves) will, in turn, improve modeling and hence will sharpen extraction of neutrino properties, in a virtuous circle.
There may be surprises, too--- current data allow for neutrino properties outside of the standard three-flavor picture, and beyond-the-Standard-Model phenomenology could also affect the supernova neutrino burst observables.

This review aims to survey how some of the neutrino mass unknowns can be determined by a supernova burst observation, with main focus on the mass ordering, also known as the mass hierarchy.    Section~\ref{sec:unknowns} briefly describes the unknowns in neutrino physics.  Section~\ref{sec:supernova-nus} describes the nature of the supernova neutrino signal.   Section~\ref{sec:flavor_transitions} describes the nature of relevant flavor transitions that will occur for supernova neutrinos. Section~\ref{sec:detection} summarizes relevant detector sensitivity and instances of detectors.  Section~\ref{sec:physics} gives examples of mass ordering signatures from a supernova burst and comments on their robustness and observability.
Section~\ref{sec:summary} is a summary.

\section{Neutrino unknowns}\label{sec:unknowns}

Thanks to experimental measurements of neutrino flavor transitions over the past few decades using diverse detectors and sources, we now
have a concise and robust model of neutrinos describing a wide array of data very well~\cite{deGouvea:2013onf, Gonzalez-Garcia:2015qrr, Olive:2016xmw}. The three-flavor neutrino model comprises three massive neutrino states connected to three flavor states by a $3\times3$ unitary mixing matrix,  $\ket{\nu_f} = \sum_{i=1}^{N} U_{fi}^* \ket{\nu_i}$, where
\begin{equation}
U= 
\left(
\begin{array}{ccc}
1 & 0 & 0\\ 
0 & c_{23} & s_{23} \\  
0 & -s_{23} & c_{23} 
\end{array} \right)
\left(
\begin{array}{ccc}
c_{13} & 0 & s_{13}e^{-i\delta}\\ 
0 & 1 & 0 \\  
-s_{13}e^{i\delta} & 0 & c_{13} 
\end{array} \right)
\left(
\begin{array}{ccc}
c_{12} & s_{12} & 0\\ 
-s_{12} & c_{12} & 0 \\  
0 & 0 & 1 
\end{array} \right), \label{eqn:mns}
\end{equation} 

$s_{ij}$ is sine of the mixing angle $\theta_{ij}$ and $c_{ij}$ is the
cosine of it. The parameters of nature in this picture are: the three mixing angles $\theta_{23}$, $\theta_{12}$ and $\theta_{13}$ plus a complex phase $\delta$ associated with CP-violating observables, as well as the three masses $m_1$, $m_2$ and $m_3$.  The mass state information is available from oscillation experiments as mass-squared differences, $\Delta m^2_{ij} \equiv m_i^2-m_j^2$;  three masses can equivalently be reported as two mass-squared differences and an absolute mass scale.

Table~\ref{tab:params} summarizes our knowledge of the mixing parameters from the global fit described in~\cite{Gonzalez-Garcia:2015qrr}.
While improved precision on all neutrino mixing parameters will be welcome, and we expect oscillation experiments to make progress in the next few decades,
there are still two quantities in this picture that are largely unknown, although there does exist at the current time some statistically-weak information about them from combined beam and reactor data.  The first unknown is the so-called `mass ordering' (MO) or `mass hierarchy', equivalent to the signs of the mass differences 
\footnote{Following recently favored usage, this review will use `mass ordering', as the word `hierarchy' suggests that some masses may be much larger than others on an
absolute scale, which may not be the case--- the masses may in fact be quasi-degenerate if their differences are much smaller than the absolute scale.}.
For `normal mass ordering' (NMO)\footnote{Because in the supernova neutrino literature, `NO' sometimes means `no oscillations', we abbreviate normal ordering as normal mass ordering, NMO, and correspondingly we abbreviate inverted mass ordering as IMO.}, we have $m_3>>m_2, m_1$, or two light and one heavy state.  For `inverted ordering' (IMO), we have  $m_2, m_1>>m_3$.   
We denote $\Delta m^2_{3\ell}$ as the larger mass-squared difference, with $\ell$=1 for NMO and $\ell=2$ for IMO.  The overall absolute mass scale is also unknown (although it is known to be less than a few eV/$c^2$), but this parameter cannot be addressed by oscillation experiments.

Another quantity largely unknown at the current time is the $\delta$ parameter associated with CP-violating observables.  However it will be very difficult to get information about this parameter from a supernova burst observation~\cite{Akhmedov:2002zj,Balantekin:2007es}.

\begin{table}[h]
 \caption{\label{tab:params} Three-flavor neutrino oscillation parameter status, from~\cite{Gonzalez-Garcia:2015qrr}.  The rightmost column indicates the primary classes of neutrino experiments on which the information is based and from which we expect future improvements.}
\centering
\begin{tabular}{|c|c|c|}

\hline
Parameter & Value, 3$\sigma$ range, any MO & Experimental information\\ \hline \hline

$\theta_{12} (^\circ)$ & 31.29 $\rightarrow$ 35.91 & Solar, reactor\\ \hline
$\theta_{23} (^\circ)$ & 38.3 $\rightarrow$ 53.3 & Atmospheric, beam\\ \hline
$\theta_{13} (^\circ)$ &7.87 $\rightarrow$ 9.11 & Reactor, beam\\ \hline
$\delta$ & 0 $\rightarrow$ 360 & Beam\\ \hline
$\Delta m^2_{21}$ (eV$^2)$ & (7.02 $\rightarrow$ 8.09) $\times 10^{-5}$ &  Solar, reactor\\ \hline
$\Delta m^2_{3\ell}$ (eV$^2)$ & (2.325 $\rightarrow$ 2.599) $\times 10^{-3}$(NMO) & Multiple, \\ 
& (-2.590 $\rightarrow$ -2.307) $\times 10^{-3}$(IMO) & including supernova \\ \hline \hline

\end{tabular}
\end{table}

There are multiple ways of going after the mass ordering experimentally.  All approaches are challenging.   A straightforward way, which will very likely succeed given sufficient exposure, is to look at neutrino and antineutrino muon to electron flavor transitions in long-baseline beam experiments.  T2K~\cite{Abe:2016tez} and NOvA~\cite{Patterson:2012zs} will give early information; we will probably need to wait for Hyper-Kamiokande~\cite{Abe:2015zbg} and the Deep Underground Neutrino Experiment~\cite{Acciarri:2015uup} for 5$\sigma$ answers. 
Similar information may be available from atmospheric neutrinos, using the naturally wide range of baselines and energies (e.g., \cite{Aartsen:2014oha,Abe:2011ts,Hofestadt:2017jfu}). Another approach is to look for subtle spectral modulations in reactor neutrino spectra as planned by JUNO~\cite{An:2015jdp}.

A core-collapse supernova burst observation is a  `method of opportunity', which, with good luck, could yield knowledge of the mass ordering before any of these experiments.  There is some model dependence, but relatively model-independent signatures do exist.   And of course, if the terrestrial experiments give us the answer first, there will be better constraints on the astrophysics.  The aim of this review is to survey some of the more robust signatures and their observability in realistic detectors.

\section{Neutrino emission from core-collapse supernovae}\label{sec:supernova-nus}

Supernovae are highly energetic and disruptive stellar outbursts.  They are understood to occur via two primary physical mechanisms.  Thermonuclear supernovae,  observationally tagged as Type Ia, are thought to be due to a thermonuclear explosion ignited after mass is accreted onto one of the stars in a binary system, although the exact mechanism is not well understood.  These events are not likely to produce very many neutrinos, although they are expected to produce some--- see section~\ref{sec:type1a}. The other main supernova type, the core-collapse supernova, corresponding observationally to Types II, Ib, Ic and some others, results from the collapse of a massive star which can no longer support its mass via nuclear burning.  These astrophysical events are well known to be generous in their neutrino production--- for a brief time, the neutrino production outshines the photon luminosity by orders of magnitude.

The physics of core collapse is the subject of supercomputer simulation studies by several groups worldwide  (see References~\cite{Mezzacappa:2005ju, Janka:2006fh, Raffelt:2012kt,Janka:2012wk, Mirizzi:2015eza} for reviews), and understanding has become more and more sophisticated over the past few decades.  Although full understanding of all details of the physical mechanisms of the collapse and subsequent explosion has not yet been achieved, the general mechanism of neutrino production is understood, and well confirmed with the observation of SN1987A.  In broad brush, the gravitational binding energy of the highly-compact remnant leaks away from the star in the form of neutrinos, thanks to the weakness of neutrino interactions in matter.  The timescale of energy loss, a few tens of seconds, is that of the trapping of the neutrinos and is set by the scale of the weak interaction with matter.

Some other general features of the neutrino production are also reasonably well understood.
The following stages of the supernova and their neutrino-producing processes in the supernova appear in most models.

\begin{itemize}

\item Infall:  as the core falls inward, there is an initial uptick of $\nu_e$ production as electrons and protons combine to form neutrons, according to $e^- + p \rightarrow n + \nu_e$.    After some milliseconds, the neutrinos become trapped in ultra-dense matter, which corresponds to a small notch in the luminosity as a function of time.

\item Neutronization burst:  after the density of matter is squeezed to its point of `maximum scrunch', the core rebounds.
The details of the process depend on the equation of state of nuclear matter.  A shock wave is formed, and as it heats the overlying matter and propagates outward, neutrinos are released.  The initial neutrino release occurs as a sharp `neutronization' (or  `deleptonization' or `breakout') burst,  highly enriched in $\nu_e$ flavor, but other flavors begin to turn on around this time.  The neutronization burst can last a few tens of ms and the luminosity has a characteristic shape as a function of time~\cite{Wallace:2015xma}.

\item Explosion and accretion: following the neutronization burst, the next few hundred milliseconds is the critical phase that determines whether the star will actually blow up, or recollapse and form a black hole.  The shock may stall, but in many models, the neutrinos themselves deposit enough energy into the envelope to reenergize the shock.  At this stage can also be seen the SASI (standing accretion shock instability), a type of `sloshing' oscillation which can manifest itself in the neutrino flux as a $\lsim$100~Hz modulation.  There can be varied structure in the neutrino flux and spectra as a function of time, depending on the details of matter accretion onto the core.  During this phase,  $\nu_e$ still tend to dominate the luminosity, but $\bar{\nu}_e$ and $\nu_x$\footnote{Because in the supernova, and also from the point of view of detection, $\nu_\mu$, $\bar{\nu}_\mu$, $\nu_\tau$ and $\bar{\nu}_\tau$ flavors are practically indistinguishable, they will be referred to collectively as `$\nu_x$', as is
conventional in the literature.} flavor components are all significant.  This stage can last up to a second or two after core bounce.

\item Cooling: this stage lasts a few tens of seconds and represents the bulk of the neutrino emission as the proto-neutron star sheds its energy via production of neutrino-antineutrino pairs of all flavors.  As a general feature, $\nu_x$ energies are greater than $\bar{\nu}_e$ energies, which are in turn greater than $\nu_e$ energies, due to increasing opacities for each; the greater the opacity, the larger the neutrinosphere radius and hence the lower the temperature at which the neutrinos decouple.  Energies gradually decrease and become more degenerate between flavors over the cooling phase.

\end{itemize}

The neutrino spectrum at a given time can be reasonably well approximated~\cite{Tamborra:2012ac} for each flavor by the following `pinched-thermal' functional form:
\begin{equation}
       \label{eq:pinched}
       \phi(E_{\nu}) = N_0 \frac{(\alpha+1)^{\alpha+1}}{\langle E_\nu \rangle \Gamma(\alpha+1)}
       \left(\frac{E_{\nu}}{\langle E_{\nu} \rangle}\right)^{\alpha} \exp\left[-\left(\alpha + 1\right)\frac{E_{\nu}}{\langle E_{\nu} \rangle}\right] \ ,
\end{equation}
where $E_{\nu}$ is the neutrino energy, $\langle E_\nu \rangle$ is the
mean neutrino energy, $\alpha$ is a `pinching parameter' (with large value associated with suppression of tails), $\Gamma$ is the  gamma function, and
$N_0$ is the total number of neutrinos emitted.  The entire flavor-time evolution of the emitted fluxes can be efficiently described by specifying the three parameters, $L$, $\langle E_{\nu} \rangle$, and $\alpha$ as a function of time for each of $\nu_e$, $\bar{\nu}_e$ and $\nu_x$.

\begin{figure}[!htbp]
\centering
\centerline{\includegraphics[width=10cm]{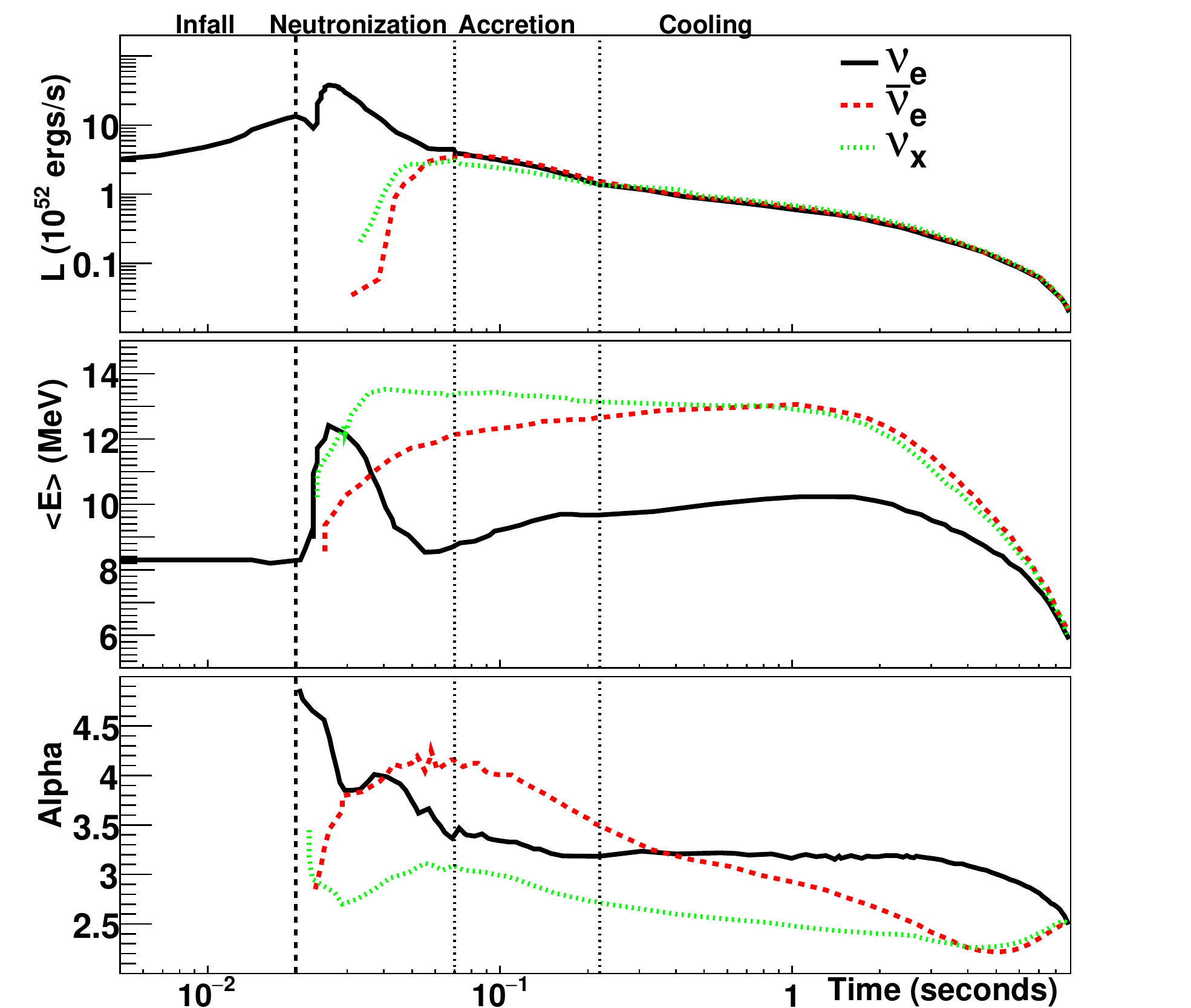}}
\caption{  Figure and caption from~\cite{Acciarri:2015uup} : Expected
  time-dependent signal for a specific flux model for an
  electron-capture supernova~\cite{Huedepohl:2009wh} at 10~kpc.  No oscillations are assumed. The
  top plot shows the luminosity as a function of time, the second plot
  shows average neutrino energy, and the third plot shows the $\alpha$
  (pinching) parameter.  The vertical dashed line at 0.02 seconds indicates
  the time of core bounce, and the vertical lines indicate different
  eras in the supernova evolution.  The leftmost time interval
  indicates the infall period.  The next interval, from core bounce to
  50~ms past core bounce, is the neutronization burst era, in which the flux is
  composed primarily of $\nu_e$.  The next period, from 50 to 200~ms past
  core bounce,
  is the accretion period. The final era, from 0.2 to 9~seconds past
core bounce, is
  the proto-neutron-star cooling period. }
\label{fig:garching}
\end{figure}

\begin{figure}[!htbp]
\centering
\centerline{\includegraphics[width=10cm]{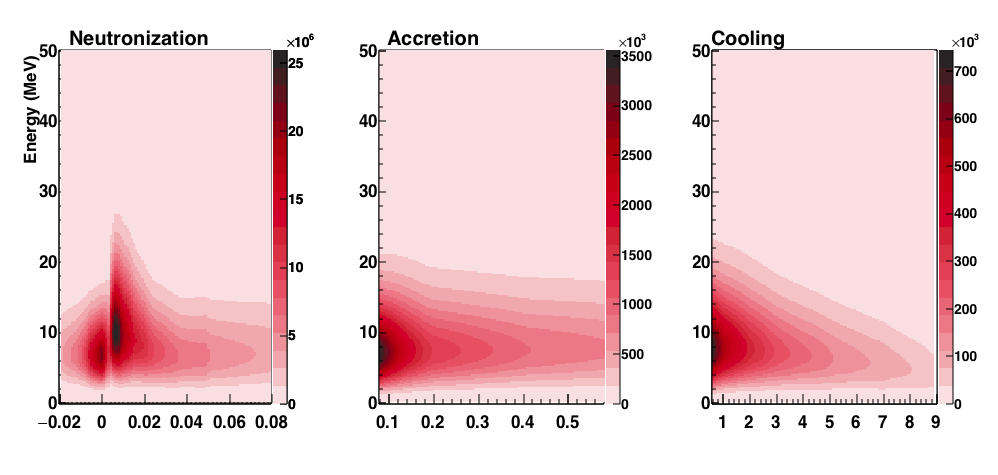}}
\centerline{\includegraphics[width=10cm]{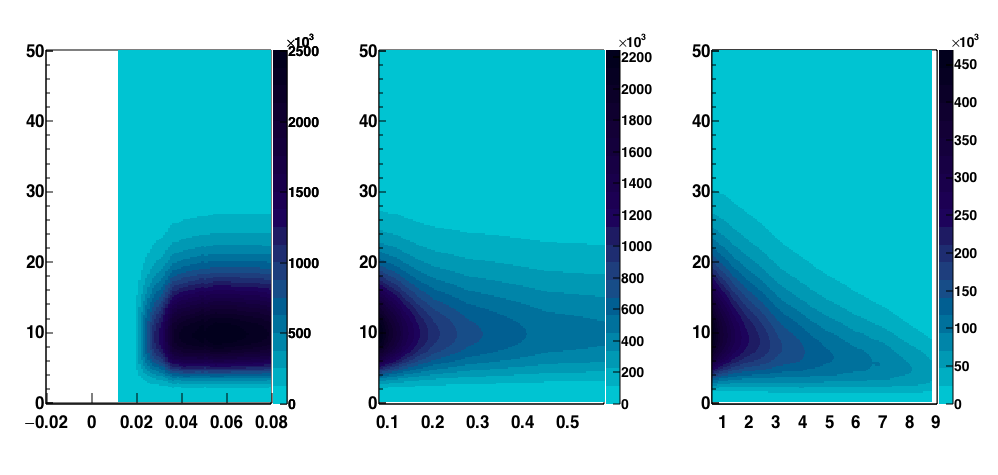}}
\centerline{\includegraphics[width=10cm]{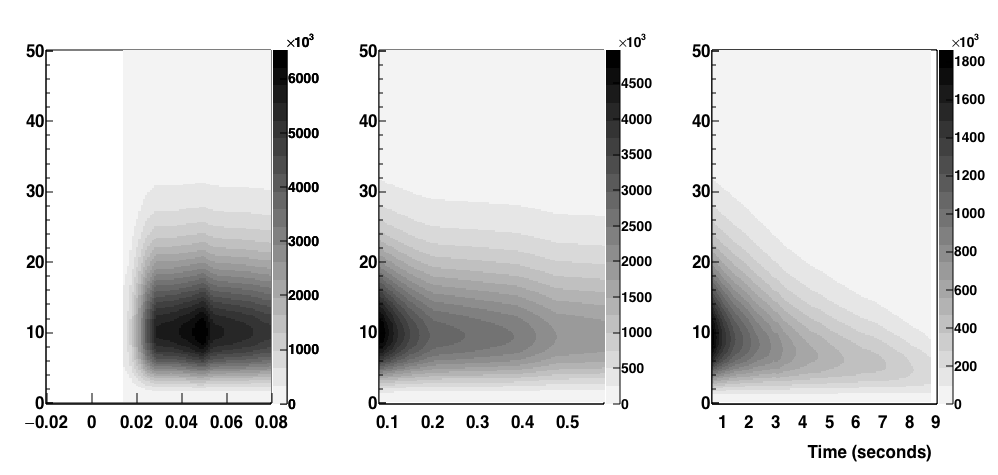}}
\caption{Example of time-dependent spectra for the electron-capture supernova model~\cite{Huedepohl:2009wh} parameterized in figure~\ref{fig:garching}, on three different timescales.   The units shown on the right-hand side of the vertical axis are neutrinos per cm$^2$ per millisecond per 0.2~MeV.  Top: $\nu_e$.  Center: $\bar{\nu}_e$.  Bottom: $\nu_x$.  Flavor transitions are not included here; note they can have dramatic effects on the spectra.  Figure from~\cite{sntutorial} (used with permission).  }
\label{fig:garching_3timescales}
\end{figure}

Figure~\ref{fig:garching} gives an example of these parameters as a function of time, describing time evolution of unoscillated neutrino fluxes for one particular model~\cite{Huedepohl:2009wh}.  Figure~\ref{fig:garching_3timescales} shows the fluxes for this model for $\nu_e$, $\bar{\nu}_e$ and $\nu_x$.   The energies and emission timescale of the few dozen neutrinos observed from 1987A in two water Cherenkov detectors (and some reported in scintillators) match this basic picture quite well.  Although most models exhibit similar features, there are variations due to different detailed assumptions in the modeling (hydrodynamics, equation of state, etc.), in addition to intrinsic variations from supernova to supernova according to properties of the progenitor and local conditions.

\section{Neutrino flavor transitions in supernovae}\label{sec:flavor_transitions}

\begin{figure}[!htbp]
\centering
\centerline{\includegraphics[width=12cm]{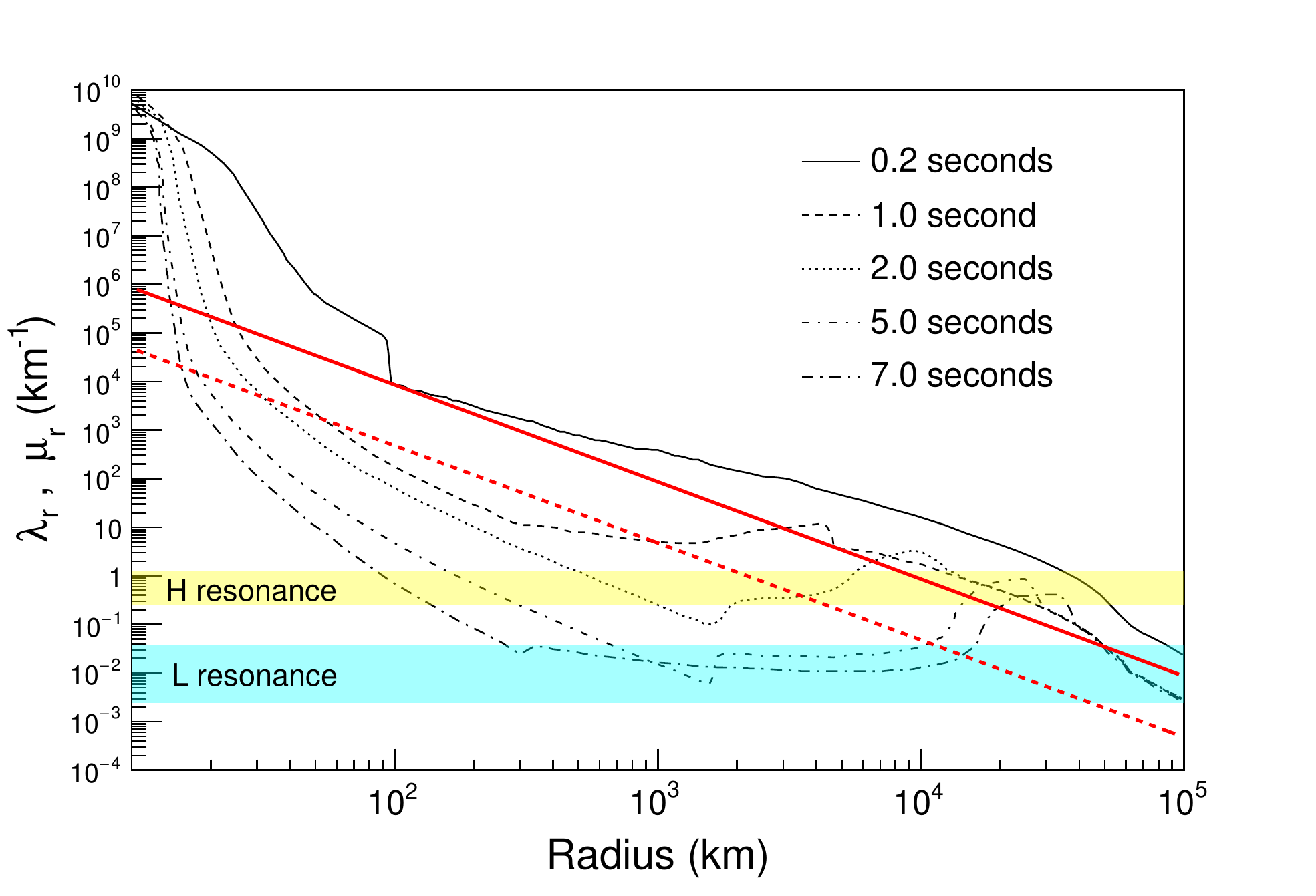}}
\caption{ Thin lines:  matter potentials $\lambda_r$ as a function of radius at various times past core bounce for a 27~$M_\odot$-progenitor core-collapse model, 
adapted from~\cite{Mirizzi:2015eza}.  Also shown as thick red lines (line style same as for matter potentials for two times) is the neutrino-neutrino potential $\mu_r$, also from~\cite{Mirizzi:2015eza}.  The horizontal bars indicate regions where matter-induced $H$ and $L$ resonances come into play. }
\label{fig:potentials}
\end{figure}

Neutrino flavor transitions are now well established experimentally, and flavor transitions driven by three-flavor mixing will certainly occur in supernovae.  
Different phenomenology holds depending on the neutrino parameters; hence, 
observed fluxes can in principle shed light on unknown neutrino parameters.
Neutrino flavor transitions in general depend on both the matter density and the flavor-dependent neutrino number densities, which
change with time as the supernova evolves.   An example of typical
expected time evolution of
these potentials as a function of radius is shown in figure~\ref{fig:potentials}.  Note
the time-dependent discontinuities associated with the shock (and reverse shock) waves that disrupt
the otherwise monotonically-decreasing matter density.  

The different types of neutrino flavor transitions relevant for supernova neutrinos are described briefly in the following subsections.\footnote{Flavor transitions due to neutrino mixing in matter will sometimes be referred to here, and are frequently referred to in the literature, as `oscillations', in spite of recent well-justified commentary~\cite{Smirnov:2016xzf} that such terminology does not appropriately discriminate adiabatic matter-induced transitions from vacuum oscillations.}.
Reference~\cite{Mirizzi:2015eza} reviews these in some detail.

\subsection{Matter effects}

When neutrinos propagate in matter, we have a regular Mikheyev-Smirnov-Wolfenstein (MSW) effect, or `matter effect'~\cite{Mikheev:1986gs,Wolfenstein:1977ue},  familiar to neutrino physicists from neutrino propagation in the Sun and Earth.   This is relatively well understood and also exhibits straightforward
mass ordering dependence.  The neutrinos feel a matter potential as a function of radial distance $r$,  $\lambda=\sqrt(2) G_F n_e(r)$, where $G_F$ is the Fermi constant and $n_e$ is the electron density.

\begin{figure}[!htbp]
\centering
\includegraphics[width=7.5cm]{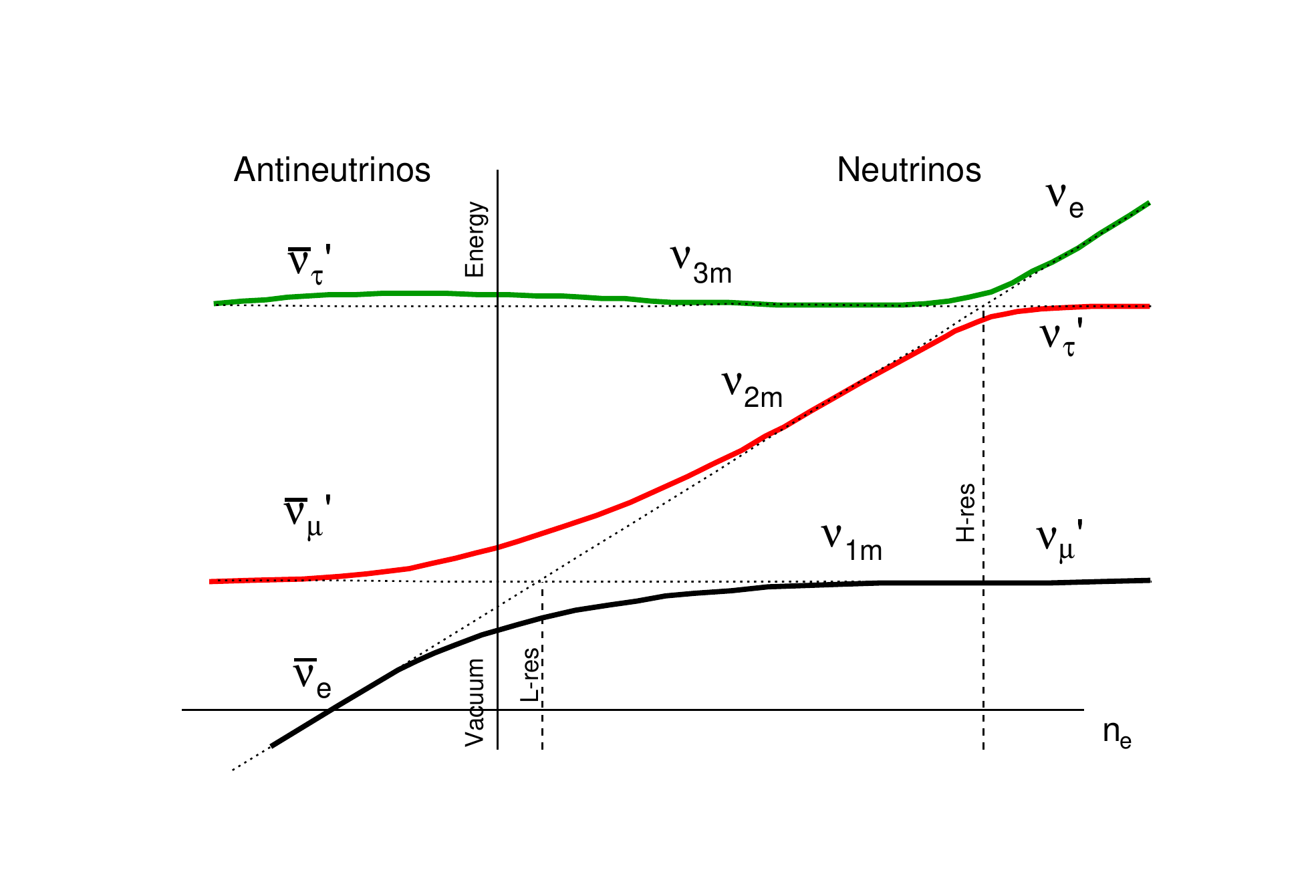}
\includegraphics[width=7.5cm]{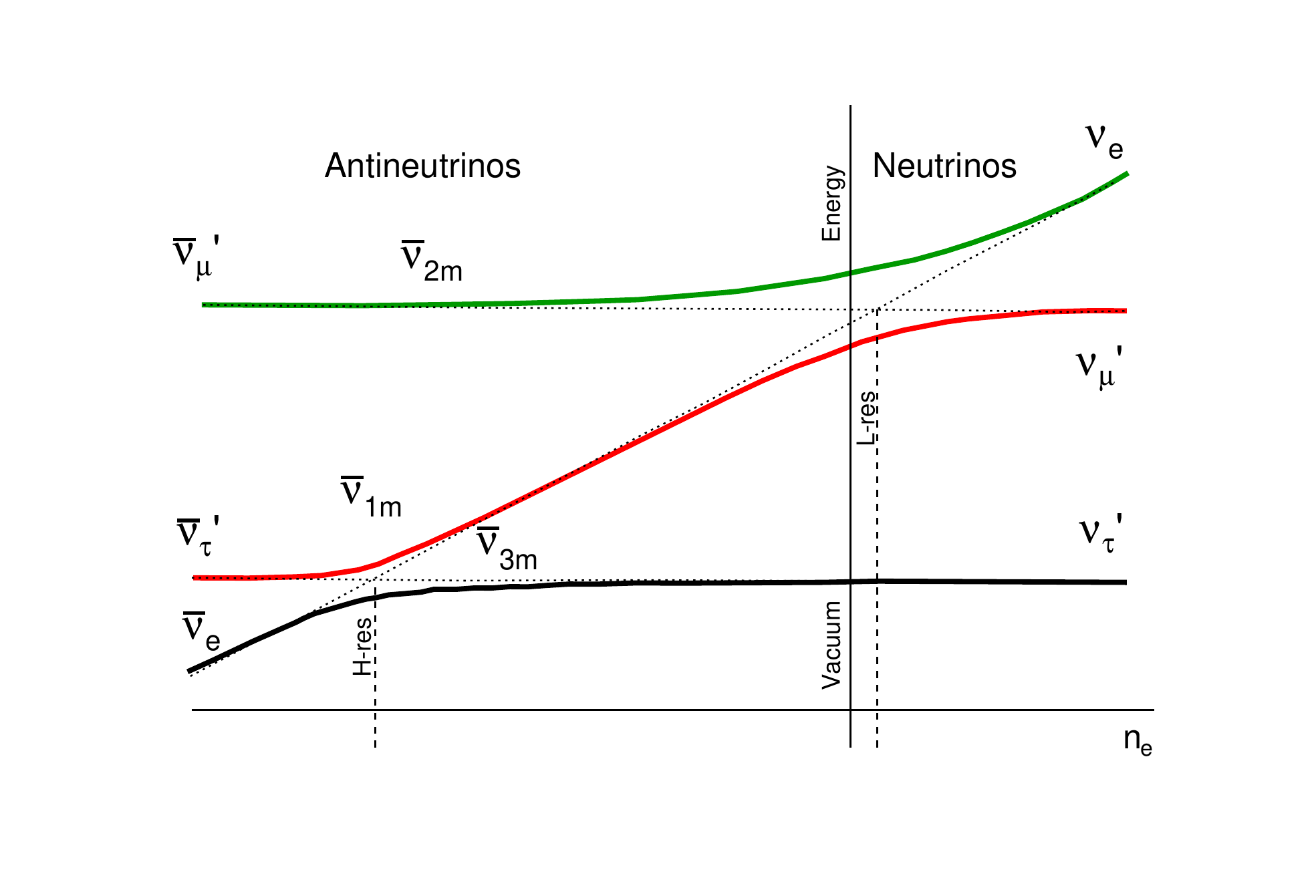}
\caption{Energy levels in matter as a function of electron density $n_e$ in a three-flavor context, following~\cite{Dighe:1999bi, Raffelt:2012kt}. The solid vertical axis indicates zero density; positive $n_e$ corresponds to neutrino states and negative $n_e$ corresponds to antineutrino states.  The states $\nu_\mu'$ and $\nu_\tau'$ represent rotations of the $\nu_\mu$ and $\nu_\tau$ states, which diagonalize the mu and tau submatrix of the effective Hamiltonian  in matter~\cite{Kuo:1986sk}.  The solid colored lines correspond to energy eigenvalues of the effective Hamiltonian in matter.     The dotted lines represent the energies of the flavor eigenstates.  The dashed vertical lines represent the $L$ and $H$ resonance values of electron density. Left: normal ordering assumption.  Right: inverted ordering assumption.}
\label{fig:msw}
\end{figure}

\subsubsection{The adiabatic case}

Figure~\ref{fig:msw} shows the neutrino eigenstate energies in matter as a function of $n_e$, for the two
mass ordering cases.   For a slowly varying density and matter potential, a neutrino born in a high-density region will propagate adiabatically as a matter eigenstate along the solid lines shown and exit the supernova in the mass eigenstate
shown by the intersection with the vacuum axis at $n_e=0$.
For antineutrinos, the potential is effectively negative, so  whereas a neutrino state
will propagate adiabatically from the right, an antineutrino initial state will
propagate adiabatically from the left.
The dotted lines of figure~\ref{fig:msw} show energies of flavor eigenstates.

At layers of specific $n_e$ where the dotted lines intersect, the neutrinos can effectively undergo resonant flavor transitions.
There are two relevant resonant matter potentials corresponding to the two mass-squared differences, $\Delta m^2_{3\ell}$ and $\Delta m^2_{12}$, on different scales; these are labeled $H$ and $L$ respectively.  Figure~\ref{fig:msw} shows
that the $H$ resonance can occur for neutrinos in the NMO case (the $H$ resonance density is on the $n_e>0$ side), and for 
antineutrinos for the IMO case (the $H$ resonance density is on the $n_e<0$ side).  The $L$ resonance occurs for neutrinos in
both MO cases (the $L$ resonance density is on the $n_e>0$ side in both cases).

Adiabatic conversion in the supernova will result in the following flavor transformations (dominated by the $H$ resonance) 
of neutrinos exiting the supernova at zero matter density:

\begin{eqnarray}  
 F_{\nu_e} &=& F^0_{\nu_x} \,\ \,\ \,\ \,\ \,\ \,\ \,\ \,\  \,\ \,\ \,\ \,\   \,\ \,\  \,\ \,\ \,\ \,\ \,\ \,\  \,\ \,\ \textrm{(NMO)} \,\ , \label{eq:msw_nmo}\\
 F_{\nu_e} &=&  \sin^2 \theta_{12} F^0_{\nu_e} +
\cos^2 \theta_{12} F^0_{\nu_x}  \,\ \,\ \,\ \,\ \textrm{(IMO)} \,\,
\label{eq:msw_imo}
\end{eqnarray} 
 and 
\begin{eqnarray}  
 F_{\bar\nu_e} &=& \cos^2 \theta_{12} F^0_{\bar\nu_e} + \sin^2 \theta_{12} F^0_{\bar\nu_x}   \,\   \,\ \,\  \,\ \,\ \,\ \,\ \,\ \,\  \,\ \,\ \textrm{(NMO)} \,\ , \label{eq:msw_nmo_anti}\\
 F_{\bar\nu_e} &=&   F^0_{\bar\nu_x}  \,\ \,\ \,\ \,\ \,\ \,\ \,\ \,\ \,\ \,\ \,\ \,\ \,\ \,\ \,\ \,\ 
 \,\ \,\ \,\ \,\ \,\ \,\ \,\ \,\ \,\ \,\ \,\ \,\ \,\  \,\
\textrm{(IMO)} \,\,\label{eq:msw_imo_anti}
\end{eqnarray}

where $F(\nu_i)$ is the flux of a given flavor ($F(\nu_x)$ represents the flux of any of either $\nu_\mu$ or $\nu_\tau$, and similarly for antineutrinos).  From these expressions, one can see that for the NMO case, the $\nu_e$ flavor component of the flux will have a spectrum (typically hotter) corresponding to that of the
original $\nu_x$ flavor; the $\bar{\nu}_e$ flux will be partially transformed.
For the IMO case, the antineutrinos will be fully transformed, and the neutrinos
will be partially transformed.  Note that in order for there to be observable effects of a flavor transition, the initial spectra for different flavors must differ sufficiently.

\subsubsection{Non-adiabatic transitions}

Neutrino propagation can occur adiabatically in a supernova, for smoothly-varying matter potentials.  However matter transitions can also occur non-adiabatically, as the matter potential can exhibit discontinuities associated with shock fronts.  If a propagating neutrino meets a matter discontinuity, a neutrino-energy-dependent level-crossing probability $P_H$ applies~\cite{Mirizzi:2015eza, Fogli:2003dw}.  The computation of this probability requires detailed knowledge of the supernova mass density profile. Since the matter discontinuity travels in space as the shock wave propagates, time- and energy-dependent signatures of the shock discontinuity can show up in the observed signal--- one could in principle see the shock propagation in the neutrino signal as a time- and
energy-dependent flavor content modulation.

We note that stochastic matter fluctuations (random inhomogeneities in
the ejecta, which are entirely plausible in a supernova) 
may wash out some of these effects. These effects are the subject of a number of
recent studies (e.g.~\cite{Friedland:2006ta, Kneller:2010sc,Lund:2013uta,Kneller:2013ska}).  Matter effects interplay as well with
self-induced flavor transitions, described in Sec.~\ref{sec:selfinduced}.

\subsubsection{Earth matter effects}\label{sec:earthmattereffects}

The neutrinos propagate as mass states after exiting the supernova, and when they arrive at Earth they have one more chance for flavor transformation if they propagate any distance in the Earth's matter.  Matter effects as the neutrinos traverse the Earth will modulate the flavor content as a function of energy.  The effect is small, but observable in large, high-energy-resolution detectors.

Under the assumption of large $\theta_{13}$ (which we know to be the true case),  the neutrino fluxes at Earth  for NMO given by~\cite{Dighe:1999bi}:
\begin{equation} 
F_{\bar\nu_e}^{\oplus} =(1 - \bar{P}_{2e}) F^0_{\bar\nu_e}+
\bar{P}_{2e} F^0_{\bar\nu_x} \,\,\,\,\,\,\,\,\,\,\,\, \textrm{and} \,\,\,\,\,\,\,\,\,\,\,\, 
F_{\nu_e}^{\oplus}=F^0_{\nu_x} \,\ ,
\label{eq:mswearth_nmo}  
\end{equation}  
and for IMO they are given by
\begin{equation} 
F_{\bar\nu_e}^{\oplus} =  F^0_{\bar\nu_x} \,\,\,\,\,\,\,\,\,\,\,\, \textrm{and} \,\,\,\,\,\,\,\,\,\,\,\,
F_{\nu_e}^{\oplus}  = (1-P_{2e}) F^0_{\nu_e}  +  P_{2e} F^0_{\nu_x} \,\ , 
\label{eq:mswearth_imo}
\end{equation} 
where $F_{\nu}^{\oplus}$ indicates the flux of neutrinos after traversing Earth matter.

The transition probabilities $P_{2e}$ and $\bar{P}_{2e}$ can be calculated~\cite{Dighe:2003vm} assuming a simplified model of Earth matter.
For a baseline distance $L$ through the mantle,  the approximate probabilities can be approximated as~\cite{Dighe:1999bi,Lunardini:2001pb}:
\begin{eqnarray}  
P_{2e} & = & \sin^2\theta_{12} + \sin2\theta^m_{12} \, \label{P2e}
 \sin(2\theta^m_{12}-2\theta_{12})  
\sin^2\left(  
\frac{\delta m^2 \sin2\theta_{12}}{4 E \,\sin2\theta^m_{12}}\,L  
\right)\,, 
\\
\bar{P}_{2e} & = & \sin^2\theta_{12} + \sin2\bar\theta^m_{12} \, \label{Pbar2e}  
 \sin(2\bar\theta^m_{12}-2\theta_{12})  
\sin^2\left(  
\frac{\delta m^2\,\sin2\theta_{12}}{4 E \,\sin2\bar\theta^m_{12}}\,L  
\right)\,,
\end{eqnarray}  
where $\theta^m_{12}$ ($\bar\theta^m_{12}$) are the effective
values of $\theta_{12}$ in the Earth matter for neutrinos (antineutrinos)
~\cite{Fogli:2001pm}.

This effect will also have some mass ordering dependence, and the prospects for observability are discussed in section~\ref{sec:earthmatter}.

\subsection{Self-induced flavor transitions}\label{sec:selfinduced}
Exotic flavor effects can occur where the
neutrino density is high enough that the potential due to neutrino-neutrino interactions cannot be ignored.
The phenomenology of these nonlinear effects is rich, and there is an extensive literature on this subject: see ~\cite{Duan:2010bg, Mirizzi:2015eza} for reviews and a more complete set of references.
It is fair to say that this is 
an area of intense and exciting theoretical study, but it has not yet converged to the point of providing robust and quantitative physics signatures of mass
ordering.

So called `collective effects' from pair conversions $\nu_e \bar{\nu}_e \rightarrow \nu_x \bar{\nu}_x$~\cite{Hannestad:2006nj} can occur, assuming appropriate flavor asymmetry.  The anisotropy of the neutrino flux can also matter,
given that the self-interaction potential depends on the angular factor
$1-{\bf v_q\cdot v_p}= 1-\cos\theta_{pq}$, where ${\bf v_q}$ and ${\bf v_p}$ are the interacting neutrino velocities and $\theta_{pq}$ is the angle between them.
Taking this angular dependence into account
can lead to significant effects on the fluxes (`multi-angle effects')~\cite{Duan:2007bt, Raffelt:2007yz,EstebanPretel:2007ec,Sawyer:2008zs}  but is computationally difficult.

\begin{figure}[!htbp]
\centering
\centerline{\includegraphics[width=8cm]{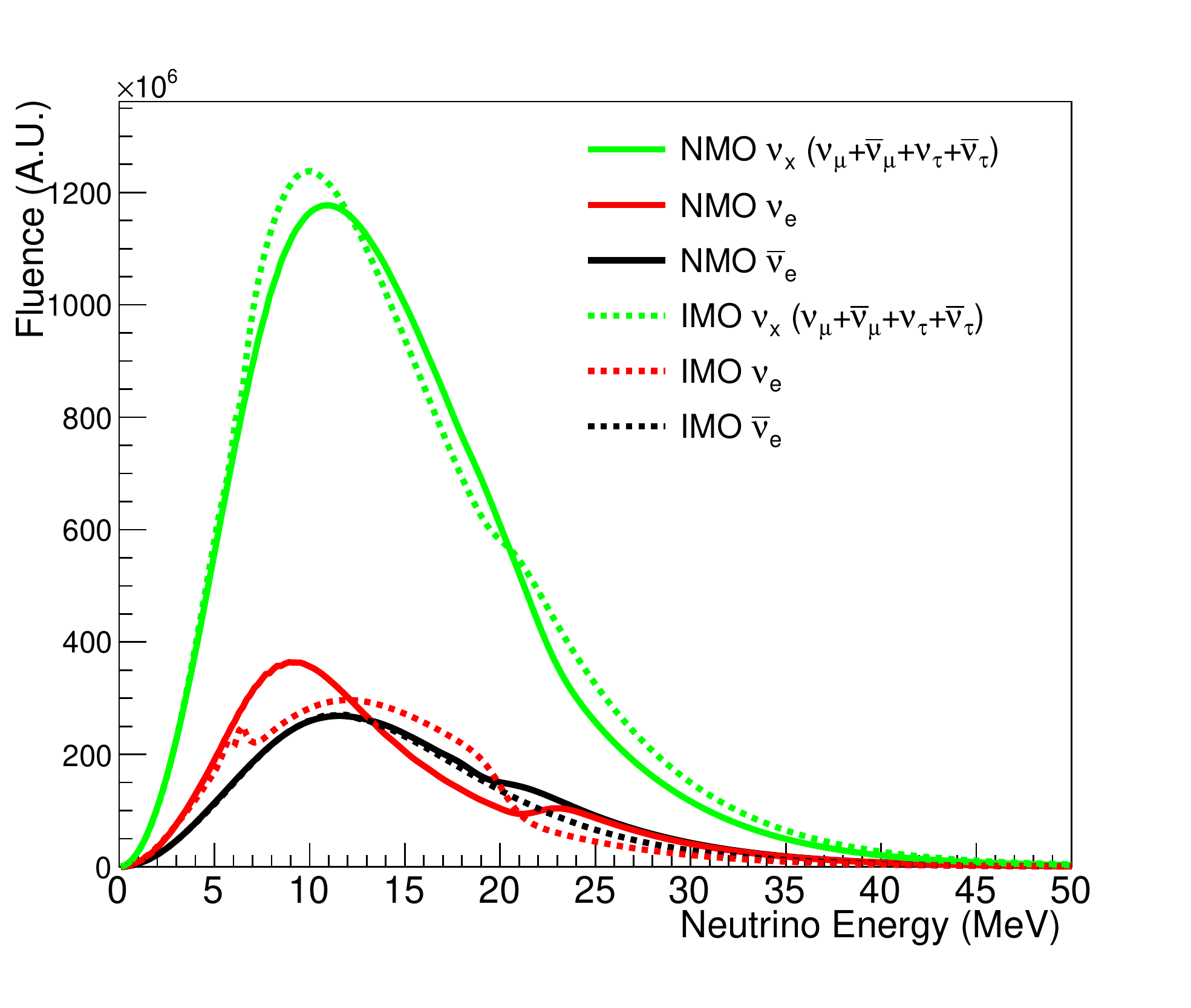}}
\caption{Example fluxes during a $\sim$1~second slice during the cooling phase for a specific model~\cite{duan, Adams:2013qkq}, showing effects of spectral swaps and splits from collective effects for NMO and IMO.    More details can be found in~\cite{Duan:2010bf}.}
\label{fig:duanfluxes}
\end{figure}

Nevertheless some likely features due to self-induced flavor transitions 
can be confidently predicted, under the assumption of certain conditions. 
Possible observable effects on the observable supernova neutrino fluxes 
include `spectral swaps', in which one flavor completely transforms into another, and`spectral splits', in which the flavor transformation occurs above or below a particular energy threshold, effectively resulting in a non-smooth spectrum with deviation from a quasi-thermal shape (see
Figure~\ref{fig:duanfluxes} for an anecdotal example).   Whether and how these
transitions occur for neutrinos or antineutrinos depends both on the mass ordering
and on the sizes of the neutrino-neutrino flavor potentials.
The presence of a large matter potential is expected to suppress self-induced
flavor transitions~\cite{EstebanPretel:2008ni}.

The neutrino-neutrino potential is 
$\mu_r=\sqrt{2}G_F \frac{F(\bar{\nu}_e)-F(\bar{\nu}_x)}{4 \pi r^2}$~\cite{Mirizzi:2015eza}, where 
$F$ are the number fluxes of the different species and $r$ is the radial distance from center of the supernova.  This potential is plotted in Fig~\ref{fig:potentials} as thick lines, and it should decrease with radius, and with time as the supernova progresses.   If this potential dominates over the matter potential $\lambda_r = \sqrt{2}G_F n_e(r)$,
where $n_e=n_{e,{\rm net}}\equiv n_{e^-}-n_{e^+}$ is the net electron density,  self-induced flavor conversions may occur.  Otherwise, matter effects will likely suppress the self-induced flavor conversions, which will likely be the
case at early times, up to the transition to the accretion phase.  

Since the conditions under which self-induced flavor transitions occur change with time, there can be complex time-dependent effects.  In the subsequent section on mass-ordering signatures, we focus on early times in the supernova evolution (about the first second, during neutronization and possibly early accretion), where it seems likely a good assumption that self-induced flavor transitions will
be a subdominant effect.

Nearly all studies of neutrino flavor transitions in supernovae so far have been done in the context of the three-flavor model.  It is worth noting that additional sterile flavors could change the phenomenology significantly~\cite{Esmaili:2014gya}.  Such possibilities will be largely ignored in this review.

\section{Neutrino detection}\label{sec:detection}

Neutrino detection and detectors are reviewed in reference~\cite{Scholberg:2012id}.  Some key points are summarized here.

\subsection{Neutrino interactions relevant for supernova neutrino detection} 

Neutrinos in the few-MeV range can interact with electrons, protons or nuclei via either charged-current (CC) or neutral-current (NC) channels.
The observables are charged or neutral products of the interactions.  They can be directly-scattered target particles, or possibly also nuclear de-excitation products (gamma rays or ejected nucleons).

Neutrinos from core collapse have energies peaking in few-tens-of-MeV range 
and only a tiny fraction have energies $>$100 MeV.  Therefore the neutrinos almost never exceed CC threshold for reactions with nuclei only for electron flavor, since thresholds for muon and tau production via CC interactions on nuclei are $\sim$100~MeV and 3.5~GeV respectively.
Therefore, while all flavors of neutrinos and antineutrinos are represented in the neutrino burst, only $\nu_e$, $\bar{\nu}_e$ and $\nu_x$ will be observable in separate channels.  Only NC interactions give access to the $\nu_x$ flavor component of the burst.

Up to now, the most important interaction experimentally has been inverse beta decay (IBD) on protons, $\bar{\nu}_e + p \rightarrow n + e^+$.  This interaction dominates for detectors with many free protons--- which includes all of the currently-running large neutrino detectors (see Sec.~\ref{sec:detectors}).  IBD not only has a relatively high cross section, but the main observable interaction product, the positron, gains an energy which tracks the neutrino energy relatively well, so measurement of its energy loss enables a $\bar{\nu}_e$ spectral measurement.  In some detectors, the neutron is captured (on free protons, or a dopant like Gd), and provides a reasonable tag of IBD, and hence of $\bar{\nu}_e$.
Another interaction of relevance, which occurs for all detectors, is elastic scattering (ES) on electrons,  $\nu+e^-\rightarrow \nu+e^-$, which proceeds via both NC and CC channels (for a supernova flux, ES interactions will be dominated by $\nu_e$ scatters).
This interaction is directional, and the directionality can be used to improve flavor tagging (and for pointing to the supernova) in some detectors.
All detectors also have some level of interaction of neutrinos and antineutrinos with nuclei, via CC channels  ($\bar{\nu}_e+(N,Z)\rightarrow (N+1,Z-1)^* +e^+$) or via NC channels ($\nu_x + (N,Z) \rightarrow \nu_x + (N,Z)^*$).  For the CC interactions, the lepton (electron or positron) is observable, and in both cases there may be observable nuclear deexcitation products, which can in principle also provide interaction channel tagging.   Typically the antineutrino interaction is suppressed in nuclei due to Pauli blocking.  A notable nuclear interaction case is $\nu_e$ on argon, $\nu_e + {}^{40}{\rm Ar} \rightarrow e^- + {}^{40}{\rm K}^*$; this reaction gives liquid argon detectors, uniquely among large supernova-neutrino-sensitive detectors,  excellent sensitivity to the electron flavor component of the flux.
NC elastic scattering on protons in scintillator and coherent elastic neutrino-nucleus scattering (CEvNS) in nuclei in dark matter detectors will also record all flavors~\cite{Horowitz:2003cz}, although detection with very large amounts of target mass is challenging.

\subsection{Supernova neutrino detectors}\label{sec:detectors}

Neutrino-matter cross sections are such that one requires a few kilotonnes of active detector mass in order to observe
$\sim$100 events for a supernova at $\sim$10~kpc.  Supernova-neutrino-sensitive detectors are also typically sited underground in order to reduce cosmogenic background, although some are on or near-surface.

Multi-kilotonne-scale neutrino detectors fall into three categories:  liquid scintillator (hydrocarbon),  liquid argon time projection chambers, and water Cherenkov (homogeneous imaging volumes or long-string photosensor detectors embedded in water or ice).  Of the Cherenkov detectors, the imaging ones are able
to do event-by-event energy and time reconstruction; in contrast, the long-string detectors map a time profile using an excess over noise of single photon hits. A few other types of supernova neutrino detectors exist, including lead-based detectors, and dark matter detectors, which are sensitive to low-energy nuclear recoils.

Of the large detector types,
water and scintillator detectors, which both have a high fraction of free protons, are primarily sensitive to $\bar{\nu}_e$ via IBD.
In contrast, liquid argon has primary sensitivity to $\nu_e$ flavor.
Other channels are observable in all detectors, and can be tagged to varying degrees, but 
are subdominant.

Table~\ref{tab:detectors} lists current and future supernova-neutrino-sensitive detectors and Tab.~\ref{tab:flavorsens} briefly summarizes flavor sensitivity.  The most promising future large detectors expected within the next decade or so 
are JUNO (scintillator)~\cite{An:2015jdp}, DUNE (liquid argon)~\cite{Acciarri:2015uup} and Hyper-Kamiokande (water)~\cite{Abe:2015zbg}.
Most of the current generation of detectors will also continue to run.

\begin{table}[h]
{\begin{tabular}{@{}cccccc@{}}%
\hline
Detector&Type &Mass (kt) &Location & Events & Status\\ \hline

Super-Kamiokande & H$_2$O& 32 & Japan& 7,000& Running\\    
LVD & C$_n$H$_{2n}$& 1 & Italy& 300& Running\\
KamLAND & C$_n$H$_{2n}$& 1 & Japan& 300& Running \\
Borexino& C$_n$H$_{2n}$& 0.3 & Italy& 100 & Running \\ 
IceCube & Long string& (600) & South Pole & ($10^6$) & Running \\  
Baksan & C$_n$H$_{2n}$  & 0.33 & Russia & 50 & Running\\  
HALO &  Pb & 0.08 & Canada & 30 & Running \\   
Daya Bay &  C$_n$H$_{2n}$ & 0.33 & China & 100 & Running \\   
NO$\nu$A$^*$ &  C$_n$H$_{2n}$ & 15  & USA &  4,000& Running \\ 
MicroBooNE$^*$ &  Ar & 0.17 & USA & 17 & Running \\  
SNO+ & C$_n$H$_{2n}$& 0.8 & Canada& 300 & Near future \\    
DUNE&  Ar & 40 &USA  & 3,000  & Future \\
Hyper-Kamiokande &  H$_2$O & 374 & Japan & 75,000 & Future \\ 
JUNO &   C$_n$H$_{2n}$& 20 &  China & 6000 & Future \\  
RENO-50 &   C$_n$H$_{2n}$& 18 &  Korea & 5400 & Future\\  
PINGU & Long string& (600) & South Pole & (10$^6$) & Future\\ 

\end{tabular}
}
\caption{Current and proposed supernova neutrino detectors as of the time of this writing.
Neutrino event 
estimates are approximate for 10~kpc; note that there is significant variation by supernova model.
An asterisk indicates a surface detector; these have more cosmogenic background.  Numbers in parentheses indicate long-string Cherenkov detectors which do not reconstruct individual interactions.}
\label{tab:detectors}
\end{table}

\begin{table}[h]
 \caption{\label{tab:flavorsens}  Summary of current and future flavor sensitivity.}
\centering
\begin{tabular}{|c|c|c|}
\hline
Flavor & Current sensitivity & Future sensitivity\\ \hline \hline

$\nu_e$ & Low (ES in SK,  &  Excellent, LAr  \\
    &            HALO) & (DUNE) \\  \hline
$\bar{\nu}_e$ & Good (SK,  & Excellent, huge \\
                       &  scintillator)   & statistics (HK, JUNO) \\ \hline
$\nu_x$   &  Low (sub-dominant  & Good \\
       & channels )               & (elastic $\nu$p scattering, \\
     &   & CEvNS) \\ \hline

\end{tabular}
\end{table}

\section{Neutrino mass physics from supernova neutrinos}\label{sec:physics}

In this section we will survey prospects for determining neutrino parameters from the supernova signal.  In some cases it is possible to quantify easily the expected sensitivity to a mass-dependent effect.  In others, however, the specific nature of the neutrino flux and spectrum is not known well enough to do this, even while the qualitative nature of the signal is generally understood.  Self-interaction effects are a particularly egregious example of this.  The reader should be assured, however, that if a signal is harvested from a Milky Way burst, physicists will be ingenious in squeezing all possible information from the data.

\subsection{Absolute mass scale}

The burst of neutrinos from a supernova bears information about the neutrino absolute mass scale, given that neutrinos have non-zero masses and hence suffer an energy-dependent time delay.  The arrival delay due to travel from distance $D$ with respect to time of arrival of a particle with velocity $c$ for a neutrino of energy $E_\nu$ and mass $m_\nu$ is

\begin{equation}
\Delta t  \sim 5.14 {~\rm ms}  \left(\frac{m_\nu}{{\rm eV}}\right)^2 
\left(\frac{10~{\rm MeV}}{E_\nu}\right)^2 \frac{D}{10 {\rm~kpc}}.
\end{equation}

At the time, $\sim$20~eV/c$^2$ neutrino mass limits based on observed time spread of the SN1987A burst neutrinos~\cite{Schramm:1990pf} were competitive with laboratory limits.   However, the current best limits from tritium beta decay endpoint experiments are now $\sim$2~eV/c$^2$~\cite{Olive:2016xmw} (and cosmology constraints are even more stringent, although model-dependent).
For few-tens-of-MeV massive neutrinos, the delays will then be less than tens of milliseconds for a travel distance of 10 kpc.   
If the neutrinos were all emitted
simultaneously, an observed neutrino event time spread could give us improved information about the absolute mass scale.  However, the emission time scale of the burst --- 10 seconds or so--- exceeds the typical delay by a large factor, so one must look for signatures of mass scale in the subtle energy-dependent timing of the arrival pattern.  The lower the energies observed, the longer the delays, so better the sensitivity.   Sensitivity has only weak dependence on distance; as the distance increases, delay increases linearly with $D$, but observed counts decrease as the inverse square of $D$. 
Large statistics, good energy resolution and low thresholds are needed.   A sharp time structure (e.g., neutrino flux cutoff due to collapse to a black hole~\cite{Beacom:2001bm}), or possibly  observation of a  gravitational wave signal of core collapse~\cite{Arnaud:2001gt,Langaeble:2016gs}  to serve as a reference time, could potentially improve sensitivity.
References~\cite{Lu:2014zma,Rossi-Torres:2015rla} estimate sensitivities of current and next-generation experiments down to some fraction of an eV.  
This is better than the current limits, but not competitive with expected next-generation experiments such as KATRIN~\cite{Drexlin:2005zt}.

\subsection{Mass ordering signatures}

This review selects a few robust signatures of mass ordering, with as little
supernova model dependence as possible.
Not emphasized here are signatures depending on neutrino self-interaction effects, due to the current partial state of understanding, although these may end up having a very important effect on the signal.

\subsubsection{The neutronization burst}\label{sec:neutronization}

\begin{figure}[!htbp]
\centering
\centerline{\includegraphics[width=10cm]{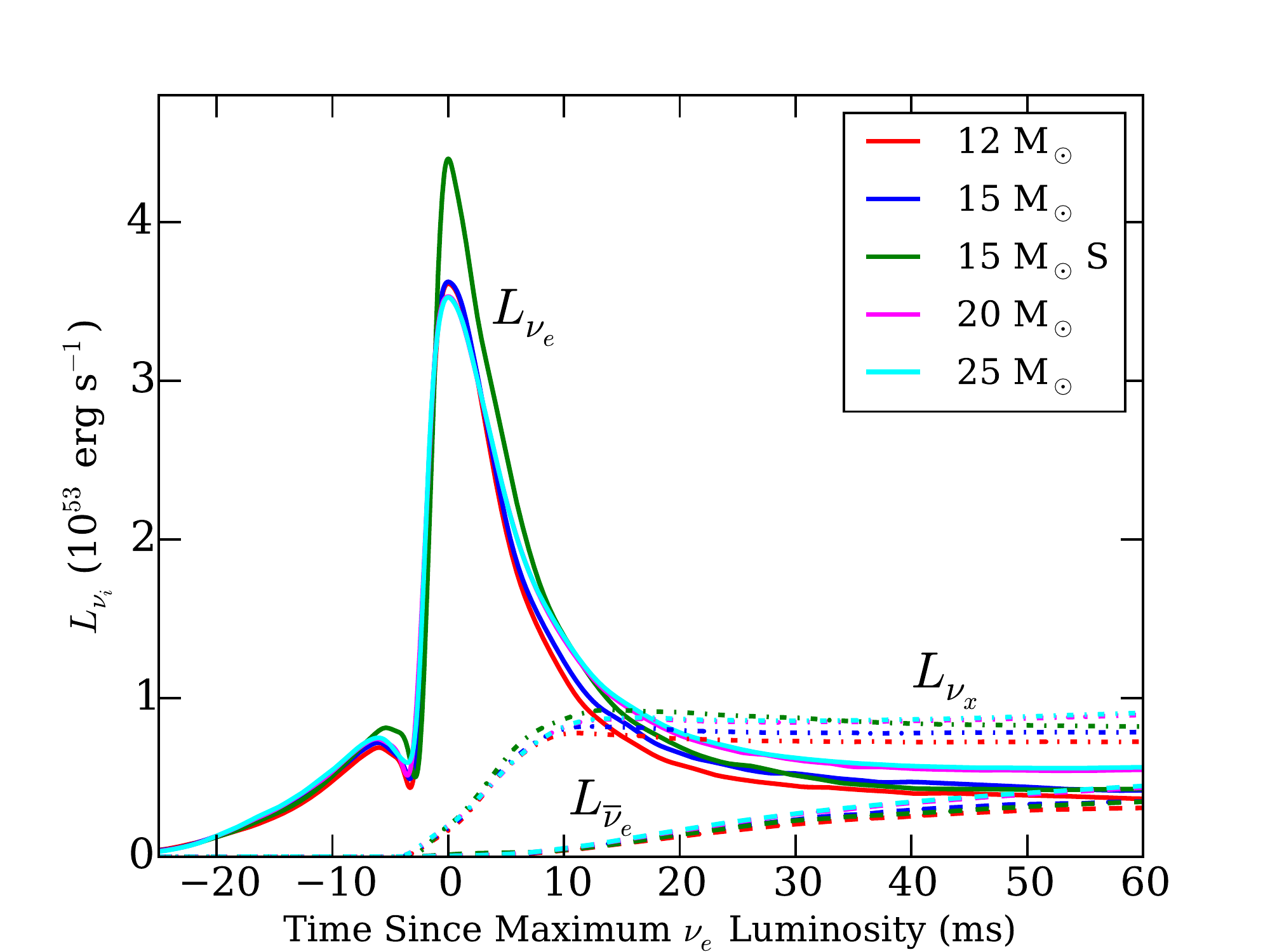}}

\caption{Figure from~\cite{Wallace:2015xma} (used with permission). Energy luminosity versus time since $\nu_e$ peak for $\nu_e$,  $\bar{\nu}_e$ and any one of $\nu_x$, for several models with different progenitors and equations of state. }
\label{fig:neutronization}
\end{figure}

Observation of the neutronization burst is probably the most robust prospect for determining the MO via a supernova burst.  The neutronization burst almost a standard candle; the time dependence of its luminosity is nearly model independent~\cite{Kachelriess:2004ds,Wallace:2015xma}: see
figure~\ref{fig:neutronization}. Its flavor is strongly dominated by $\nu_e$.  Because the electron neutrinos escape from regions for which the lepton asymmetry is such that self-interaction has a negligible effect~\cite{Hannestad:2006nj,Mirizzi:2015eza}, one expects the burst to be processed by MSW effects only, in a MO-dependent way.  This greatly simplifies the interpretation of the signal.

\begin{figure}[!htbp]
\centering
\centerline{\includegraphics[width=11cm]{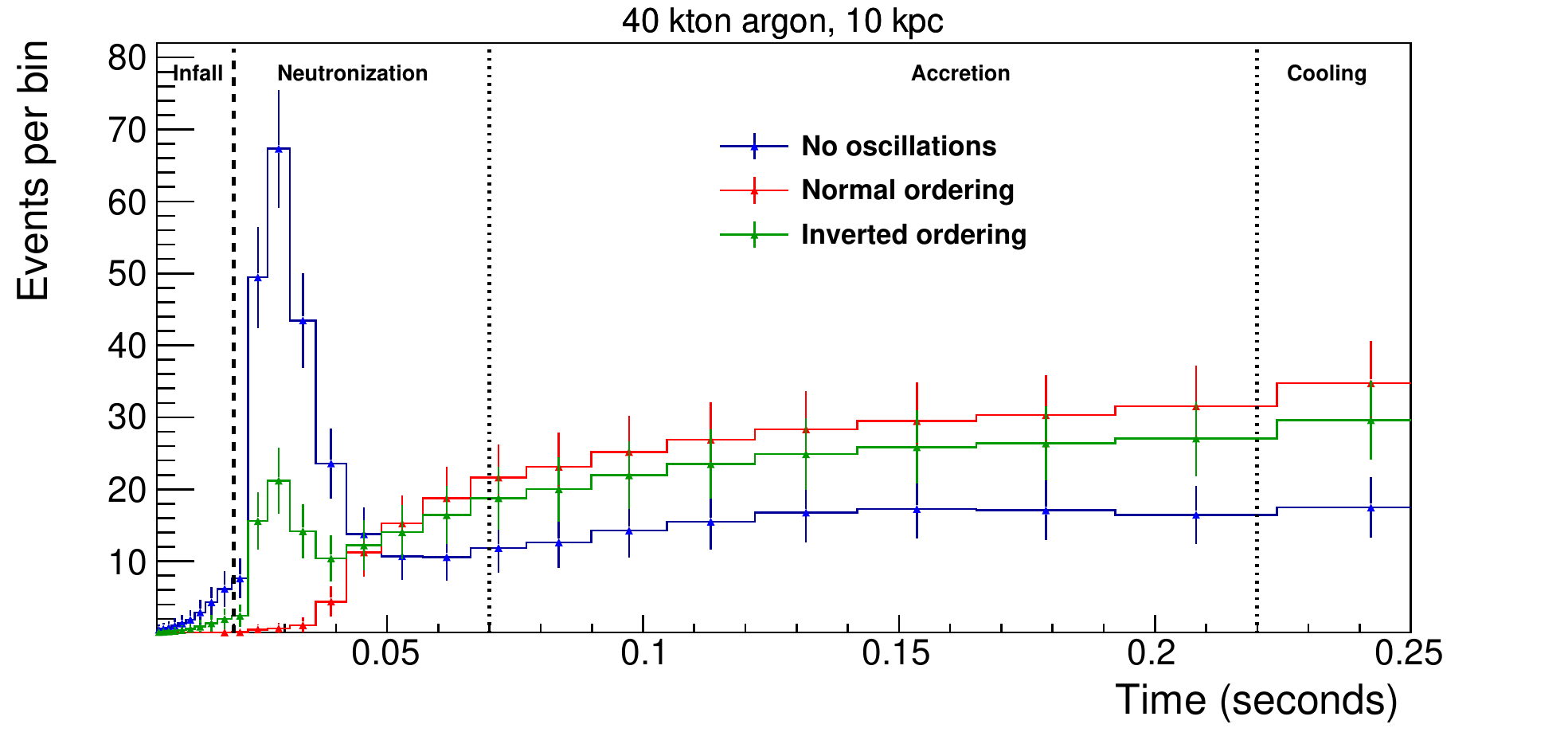}}
\centerline{\includegraphics[width=11cm]{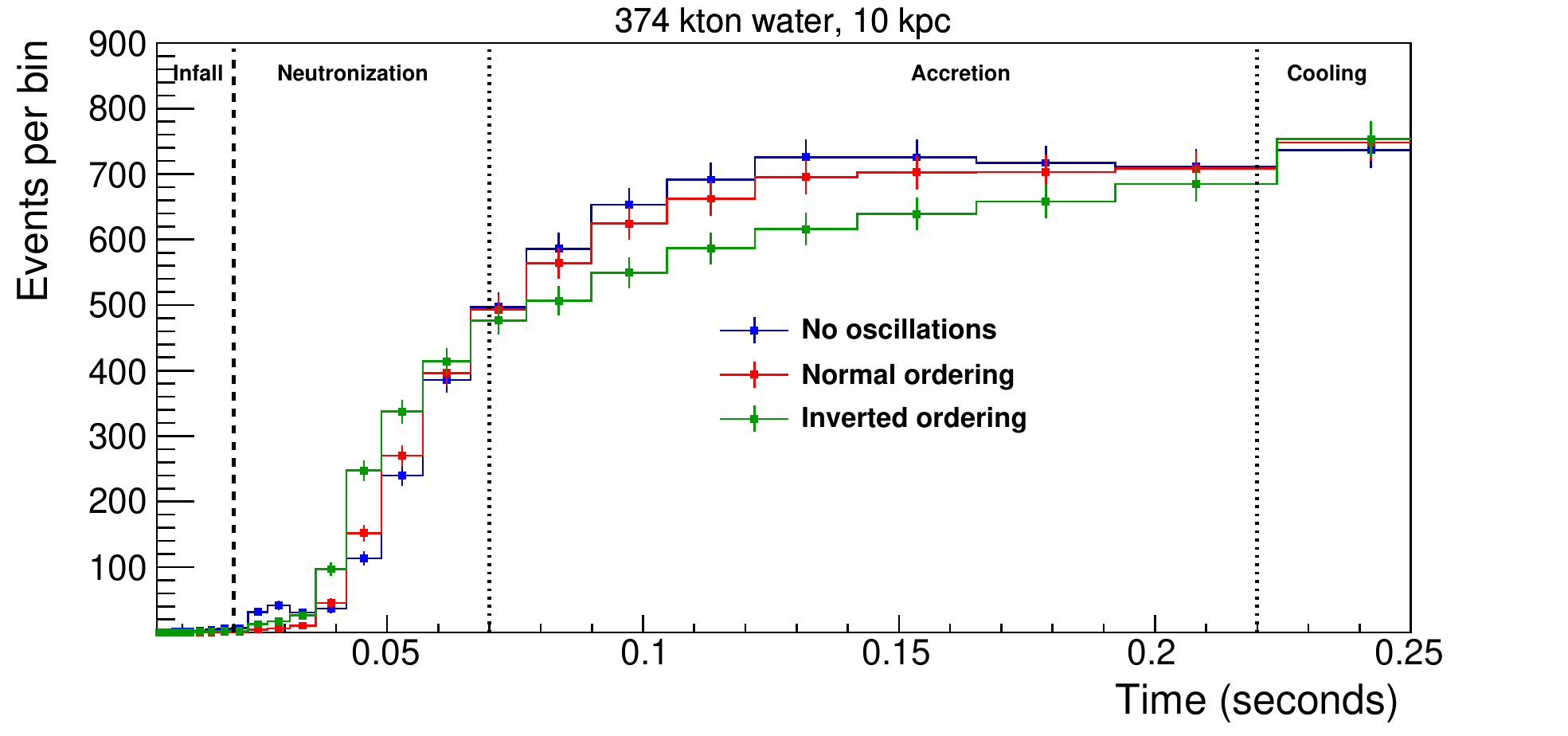}}
\centerline{\includegraphics[width=11cm]{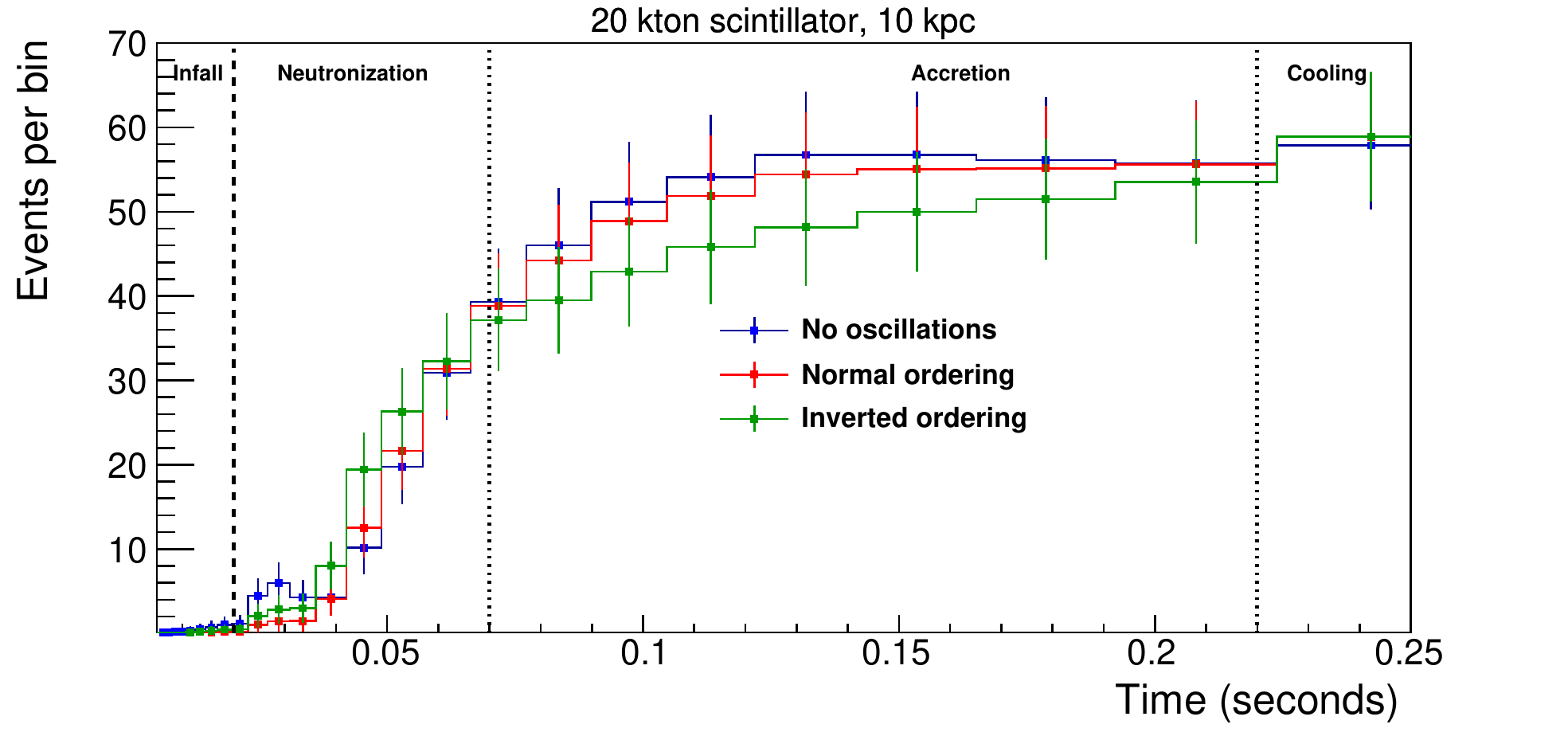}}
\caption{Expected event rates as a function of time for the electron-capture supernova model in~\cite{Huedepohl:2009wh} for 40 kilotonnes of argon (DUNE-like, top), 374 kilotonnes of water (Hyper-K-like, middle) and 20 kilotonnes of scintillator (JUNO-like, bottom), during early stages of the event--- the neutronization burst and early accretion phases, for which self-induced effects are unlikely to be important.  Shown for each are the event rate for the unrealistic case of no flavor transitions (blue), the event rate including the effect of MSW transitions for the normal (red)  and inverted (green) hierarchies.  Error bars are statistical, in unequal time bins.}
\label{fig:early_time}
\end{figure}

According to equations~\ref{eq:msw_nmo},\ref{eq:msw_imo}, the $\nu_e$ flux will be entirely swapped with $\nu_x$ for the case of NMO.  Since there is very little $\nu_x$ present for the duration of the neutronization burst, there will be very little $\nu_e$ to observe.
In contrast, 
for the case of IMO the $\nu_e$ component will be only partially swapped (see equations~\ref{eq:msw_nmo_anti},\ref{eq:msw_imo_anti}).    In other words, the neutronization burst is suppressed for IMO, but suppressed even more strongly for NMO.  Hence, the signature of NMO is an absent or highly suppressed neutronization burst in a $\nu_e$-sensitive detector.  The signature of IMO is an observable neutronization burst.  Such an interpretation could be strengthened by (non)-observation of other flavors at the time of the $\nu_e$ peak, in detectors with NC sensitivity. 

This suppression should be observable easily in a liquid argon detector, but also should be visible to some extent in a large water detector, for which $\nu_e$ can be seen via ES.
See figure~\ref{fig:early_time} for an example of the expected neutronization burst (or its absence) in large argon, water and scintillator detectors.

\subsubsection{Early time profile}\label{sec:earlytime}

We can also fairly robustly constrain the MO by including the few-hundred millisecond timescale beyond the neutronization burst and considering the overall shape of the early time profile,  as the neutronization burst transitions to the accretion era.  The flux should remain dominated by $\nu_e$ during this period.   During the early accretion era, as for the neutronization burst, flavor transitions may be dominated by MSW effects, and therefore understanding of the MO signature is relatively robust.

For the first $\sim$50 ms where measured $\nu_e$ flavor dominate, we expect the IMO to give a larger $\nu_e$ signal, as per Sec.~\ref{sec:neutronization}.  After around 60 ms, in the accretion phase,  most of the observed signal in a high-statistics detector will be $\bar{\nu}_e$.  Here, for MSW transitions, the  $\bar{\nu}_e$ will be mostly untransformed for NMO, whereas for IMO, the $\bar{\nu}_e$ will have mostly swapped with $\nu_x$, which has lower flux during
this period; hence NMO will give the larger signal.

The net outcome is the IMO gives a flatter time profile and NMO gives a more sharply rising time profile in $\bar{\nu}_e$.  A detector like IceCube, mostly sensitive to $\bar{\nu}_e$, will be able to track the time profile well enough to distinguish the shapes~\cite{evan}.

\subsubsection{Shock wave effects}\label{sec:shockwave}

As the shock wave progresses through the overlying stellar matter, the density changes discontinuously.  MSW effects will occur when the matter potential matches that required for level transition.  Since the matter potential changes with time, this can lead to flavor transition, and hence a flavor composition change, as a function of time~\cite{Schirato:2002tg, Dasgupta:2005wn, Choubey:2006aq, Kneller:2007kg}.  The specific signal will be MO-dependent.
We note that in this phase, neutrino self-interaction effects may also be affected
by the shock wave.  For example, ~\cite{Gava:2009pj} describes MO-dependent modulations of the observable signal in time and energy.

\subsubsection{Spectral swaps and splits}

\begin{figure}[!htbp]
\centering
\includegraphics[width=7cm]{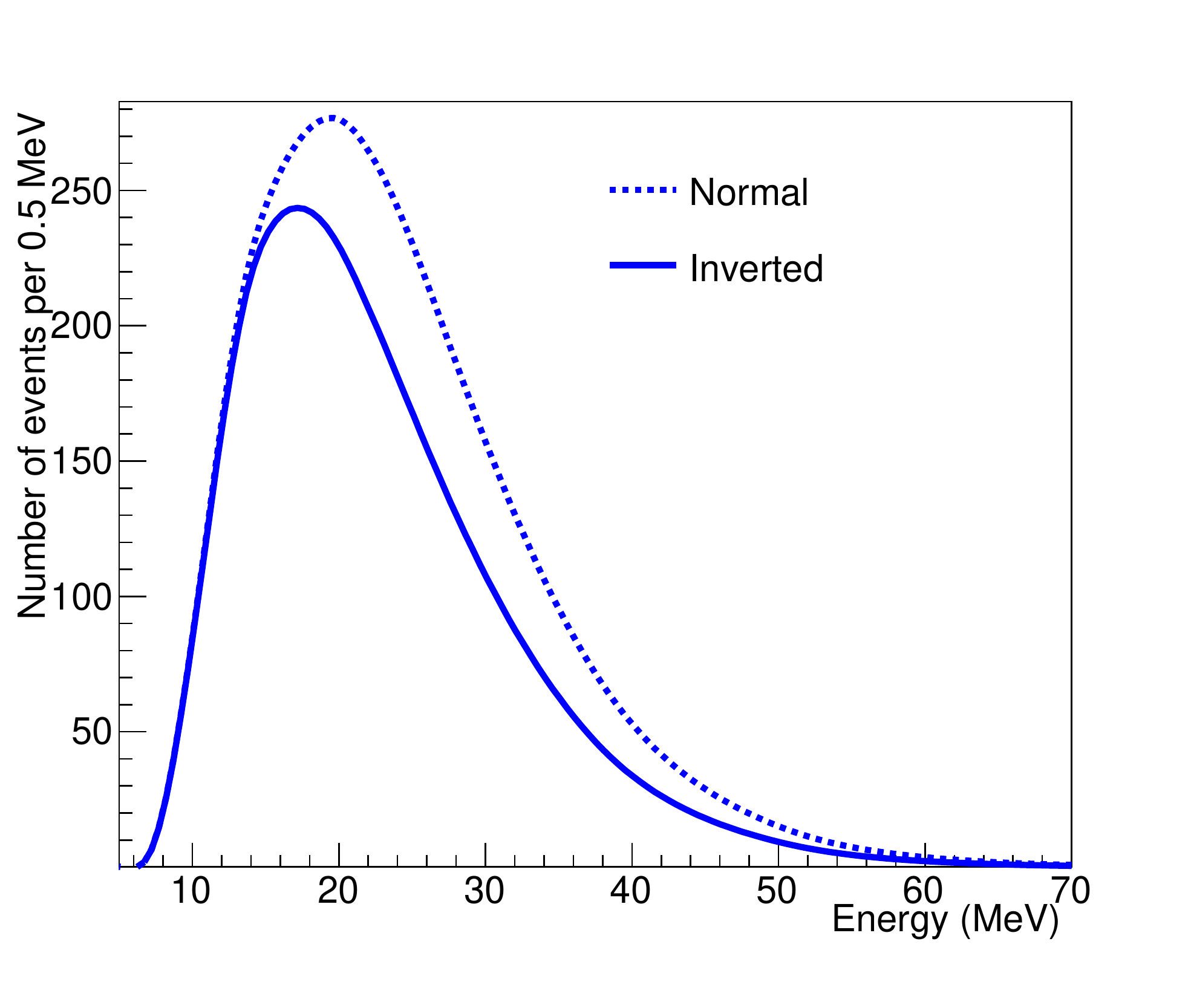}
\includegraphics[width=7cm]{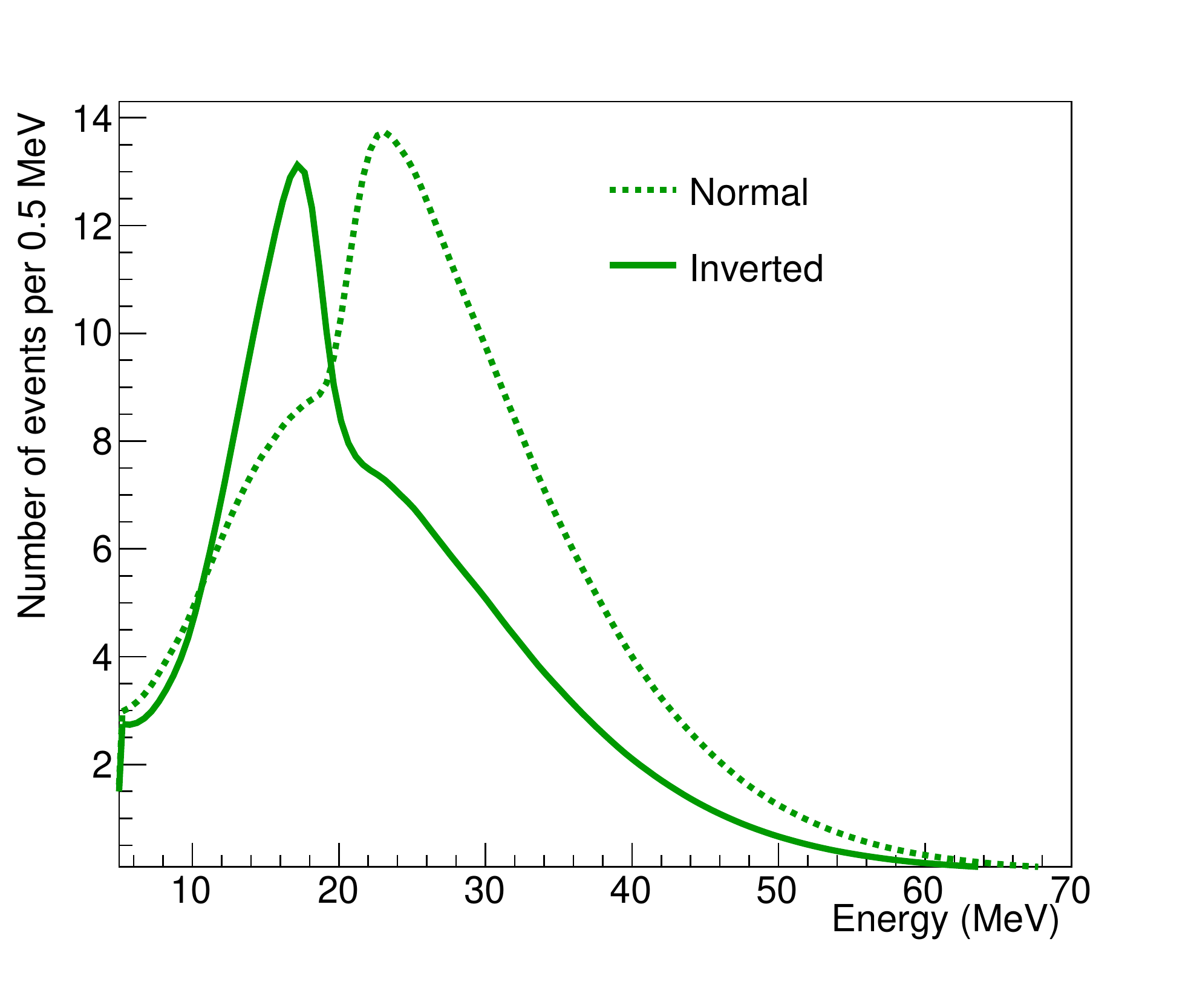}
\caption{ Estimated total measured event spectra~\cite{Adams:2013qkq}, calculated according to~\cite{snowglobes},  in 374 kilotonnes of water (dominated by $\bar{\nu}_e$)  and 40 kilotonnes of argon (dominated by $\nu_e$) in a $\sim$ 1-second cooling-phase time slice for the flux (with self-induced effects) shown in figure~\ref{fig:duanfluxes}.  There are different distinctive features in each mass ordering case. }
\label{fig:duan}
\end{figure}

A potential dramatic MO-dependent effect on the time dependent supernova neutrino spectra is the so-called `spectral split' due to collective effects~\cite{Raffelt:2007cb,Raffelt:2007xt,Dasgupta:2009mg,Dasgupta:2010cd}.  The effect is that the neutrino flavor spectra are swapped above or below a particular energy threshold, depending on both the initial relative flavor luminosities and in neutrinos or antineutrinos depending on the hierarchy.
Observationally this results in non-thermal observed spectral shapes in either neutrinos or antineutrinos~\cite{Choubey:2010up}: see figure~\ref{fig:duan} for the example observed spectra corresponding to the fluxes of figure~\ref{fig:duanfluxes}.  These non-thermal spectral distortions can also track the propagation of the shock wave.

There is enough variety of phenomenology in the literature describing this kind of signature that this signature can not yet be considered
robust.  However, there are potentially multiple signatures at different times in the star's evolution.

\subsubsection{Earth matter effect}\label{sec:earthmatter}

\begin{figure}[!htbp]
\centering
\centerline{\includegraphics[width=9cm]{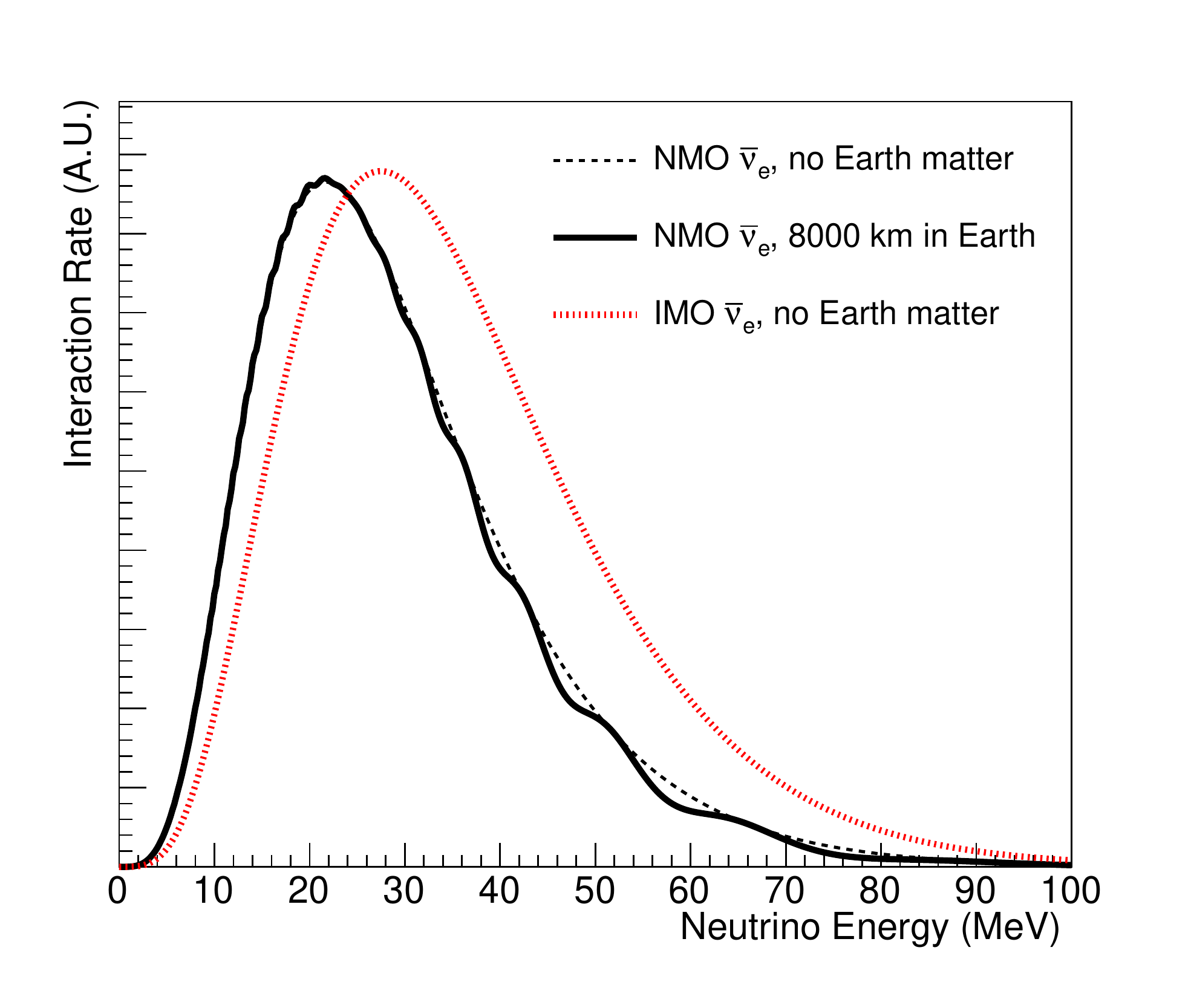}}
\caption{Matter effects in the Earth:  Expected $\bar{\nu}_e$ interaction rates ($\propto E_\nu^2 F(E_\nu)$) (\cite{Totani:1997vj}) under the assumption of MSW flavor transformation in the supernova and Earth (no self-interaction effects), for normal (dashed black) and inverted (dotted red) mass orderings. The solid black line shows the oscillatory effect of matter effects in the Earth,
for an assumed path length of 8000~km through the mantle.}
\label{fig:earthmattereffect}
\end{figure}

Finally, the neutrinos arrive at Earth as mass states and may travel through matter before arriving at a detector.  They will undergo conventional matter effects in the Earth which will modulate the spectrum according to the distance and densities traversed in matter, with flavor-dependent effect depending on the MO, as described in section~\ref{sec:earthmattereffects}.  Wiggles in the spectrum of around $\sim$few-10~MeV frequency and amplitude of a few percent will appear in antineutrinos in the NMO case, and in neutrinos in the IMO case.  Figure~\ref{fig:earthmattereffect} shows an example of the effect for $\bar{\nu}_e$ observed by IBD.  Fourier analysis of a well-measured energy spectrum could potentially identify the
ordering based on presence or absence of a peak in the appropriate channels~\cite{Dighe:2003jg,Dighe:2003vm}.

This effect is relatively well understood.  However it does require Earth shadowing~\cite{Mirizzi:2006xx} and sufficiently different primary flavor spectra in the cooling phase.   A challenge from an observational point of view is that both good energy resolution and large statistics will be required to resolve the wiggles~\cite{Borriello:2012zc}.  The best prospect is for a large scintillator detector like JUNO with excellent energy resolution~\cite{Liao:2016uis}, although for more optimistic models one could observe an Earth-matter-induced difference between signals in large water Cherenkov detectors~\cite{Dighe:2003be} with different pathlengths through the Earth's mantle.

\subsubsection{Non-core-collapse supernovae}\label{sec:type1a}

\begin{figure}[!htbp]
\centering
\centerline{\includegraphics[width=11cm]{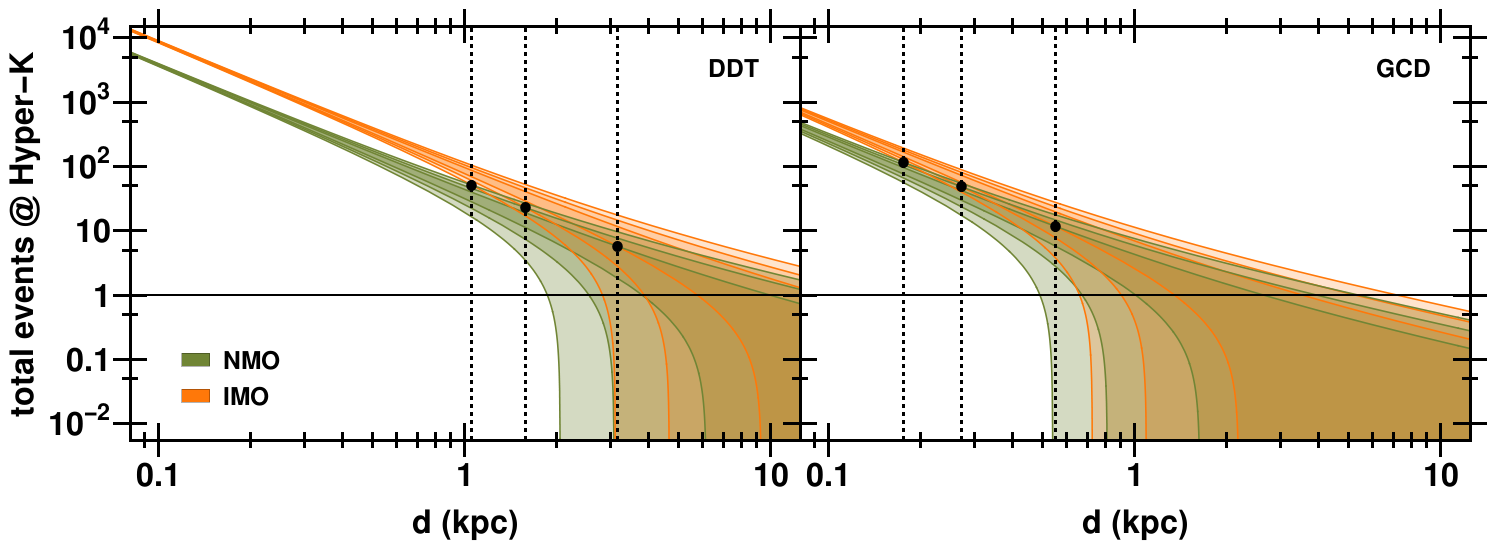}}
\caption{  From reference~\cite{Wright:2016gar} (used with permission): Expected events in  Hyper-K vs. distance to the Type Ia SN.  The green shaded regions are for the NMO assumptions,
with darkest to lightest corresponding to 1$\sigma$, 2$\sigma$ and 3$\sigma$ Poisson event count ranges.  The orange regions show the same for the IMO assumption.  
The left plot corresponds to the DDT model and the right plot corresponds to the GCD model (see text).  The black dots with dotted vertical lines show the distances at which NMO and IMO can be distinguished for a given $\sigma$ level, assuming the explosion model is known.}
\label{fig:type1a}
\end{figure}

As a final note, we mention the Type Ia (thermonuclear) supernova case.   As noted above, these have a quite different physical mechanism from the core-collapse scenario, and are expected to be much fainter in neutrinos.   The mechanism is not fully understood, and is thought to take place either according to a `deflagration-to-detonation transition' (DDT)
or a `gravitationally confined detonation' (GCD) scenario.   Still, there should be some neutrino production ~\cite{Odrzywolek:2003vn, Wright:2016xma,Wright:2016gar} albeit some orders of magnitude less than for a core-collapse event; the DDT scenario is expected to produce a fainter neutrino flux.   For the more neutrino-generous DDT model,  Hyper-K would detect a handful of events at 10 kpc, and Super-K and DUNE would see a few events at 1~kpc; the distance sensitivity is reduced to $\sim$1~kpc and $\sim$0.3~kpc respectively for the GCD model.    From the point of view of distinguishing the mass ordering, an observation of a thermonuclear supernova has the advantage that the flavor transition effects are purely MSW--- there are no self-interaction effects to confound the interpretation.  And for a nearby Type Ia supernova, we may be able to observe the event closely enough in electromagnetic channels to understand the mechanism.   According to ~\cite{Wright:2016gar}, if the mechanism is understood, with future generation detectors we should be able to distinguish the MO at 1$\sigma$ for a Type Ia at $\sim$3~kpc for DDT and $\sim$0.5~kpc for GCD: see figure~\ref{fig:type1a}.

As a final note, the neutrino signal from pair-instability supernovae has recently been explored~\cite{Wright:2017zyq}. Pair-instability supernovae are a class of less common, superluminous supernovae from very massive carbon-oxygen-core progenitors, in which a thermonuclear explosion follows collapse due to conversion of photons to electron-positron pairs.  In principle the emitted neutrinos could also exhibit observable MO-dependent effects.  However, as for Type Ia supernovae, the neutrino signal is relatively faint and observation would require a nearby explosion.

\section{Summary}\label{sec:summary}

In summary, a core-collapse supernova in our galaxy or nearby will bring tremendous information, via the flavor-energy-time profile of the neutrino flux.  Embedded in the signal will be information on neutrino properties, including on neutrino masses.  While the absolute mass scale information will not likely be competitive with next-generation terrestrial experiments, the MO information may well be.  There are multiple signatures of varying robustness summarized in table~\ref{tab:signature_summary}.  As understanding of core-collapse supernova phenomenology improves, so will the robustness of the signatures.  It is also very possible that terrestrial experiments will yield MO information first.  In this case, there is only benefit for extracting other information from the supernova, and this review highlights aspects of the phenomenology where information from neutrino experiments will help to constrain the astrophysical interpretations.

\begin{table}[h]
 \caption{\label{tab:signature_summary} Qualitative summary of supernova MO signatures.}
\begin{tabular}{|c|c|c|c|c|}

\hline
Signature & Normal &  Inverted & Robustness & Observability\\ \hline \hline

 Neutronization & \small Very & \small Suppressed & \small Excellent & \small Good, need $\nu_e$\\ 
 & \small suppressed & & & \\ \hline 
Early time profile & \small Low then & \small Flatter & \small Good & \small Good, \\ 
     & \small high &  & \small possibly some& \small especially IceCube\\ 
     &                   && \small self-interaction &\\
     &                   & &  \small effects & \\ \hline
Shock wave & \small Time- & \small Time- & \small Fair,  & \small May be \\ 
        &\small dependent & \small dependent & \small entangled with & \small statistics\\ 
       &                 &                    & \small  self-interaction     &  \small limited\\ 

  & & & \small effects& \\ \hline
   
Collective effects &\multicolumn{2}{|c|} {\small Various time- and energy-} & \small Unknown, but & \small Want all\\ 
 & \multicolumn{2}{|c|} {\small dependent signatures} & \small multiple signatures &\small flavors\\ \hline
Earth matter & \small Wiggles in $\bar{\nu}_e$ & \small Wiggles in $\nu_e$ & \small Excellent & \small Difficult, need\\ 
& & & & \small energy resolution,\\ 
& & & &  \small Earth shadowing\\  \hline
Type Ia &  \small Lower flux & \small Higher flux& \small Moderate & \small Need large detectors,  \\
  & & & & \small very close SN\\ \hline \hline

\end{tabular}
\end{table}

\normalsize
\newpage

\section*{References}

\bibliographystyle{iopart-num}
\bibliography{refs}

\providecommand{\newblock}{}
\begin{thebibliography}{10}
\expandafter\ifx\csname url\endcsname\relax
  \def\url#1{{\tt #1}}\fi
\expandafter\ifx\csname urlprefix\endcsname\relax\def\urlprefix{URL }\fi
\providecommand{\eprint}[2][]{\url{#2}}

\bibitem{Bionta:1987qt}
Bionta R~M {\em et~al.\/} 1987 {\em Phys. Rev. Lett.\/} {\bf 58} 1494

\bibitem{Hirata:1987hu}
Hirata K {\em et~al.\/} 1987 {\em Phys. Rev. Lett.\/} {\bf 58} 1490--1493

\bibitem{Alekseev:1987ej}
Alekseev E~N, Alekseeva L~N, Volchenko V~I and Krivosheina I~V 1987 {\em JETP
  Lett.\/} {\bf 45} 589--592

\bibitem{Schramm:1990pf}
Schramm D~N and Truran J~W 1990 {\em Phys. Rept.\/} {\bf 189} 89--126

\bibitem{Raffelt:1998hw}
Raffelt G~G 1999 {\em AIP Conf. Proc.\/} {\bf 490} 125--162

\bibitem{Vissani:2014doa}
Vissani F 2015 {\em J. Phys.\/} {\bf G42} 013001 (\textit{Preprint}
  \eprint{1409.4710})

\bibitem{Scholberg:2012id}
Scholberg K 2012 {\em Ann.Rev.Nucl.Part.Sci.\/} {\bf 62} 81--103
  (\textit{Preprint} \eprint{1205.6003})

\bibitem{Gonzalez-Garcia:2015qrr}
Gonzalez-Garcia M~C, Maltoni M and Schwetz T 2016 {\em Nucl. Phys.\/} {\bf
  B908} 199--217 (\textit{Preprint} \eprint{1512.06856})

\bibitem{deGouvea:2013onf}
de~Gouvea A {\em et~al.\/} (Intensity Frontier Neutrino Working Group) 2013
  {Working Group Report: Neutrinos} {\em {Proceedings, 2013 Community Summer
  Study on the Future of U.S. Particle Physics: Snowmass on the Mississippi
  (CSS2013): Minneapolis, MN, USA, July 29-August 6, 2013}\/}
  (\textit{Preprint} \eprint{1310.4340})
  \urlprefix\url{http://inspirehep.net/record/1260555/files/arXiv:1310.4340.pdf}

\bibitem{Olive:2016xmw}
Patrignani C {\em et~al.\/} (Particle Data Group) 2016 {\em Chin. Phys.\/} {\bf
  C40} 100001

\bibitem{Akhmedov:2002zj}
Akhmedov E~K, Lunardini C and Smirnov A~{\relax Yu} 2002 {\em Nucl. Phys.\/}
  {\bf B643} 339--366 (\textit{Preprint} \eprint{hep-ph/0204091})

\bibitem{Balantekin:2007es}
Balantekin A~B, Gava J and Volpe C 2008 {\em Phys. Lett.\/} {\bf B662} 396--404
  (\textit{Preprint} \eprint{0710.3112})

\bibitem{Abe:2016tez}
Abe K {\em et~al.\/} (T2K) 2016  (\textit{Preprint} \eprint{1607.08004})

\bibitem{Patterson:2012zs}
Patterson R~B (NOvA) 2012  [Nucl. Phys. Proc. Suppl.235-236,151(2013)]
  (\textit{Preprint} \eprint{1209.0716})

\bibitem{Abe:2015zbg}
Abe K {\em et~al.\/} (Hyper-Kamiokande Proto-Collaboration) 2015 {\em PTEP\/}
  {\bf 2015} 053C02 (\textit{Preprint} \eprint{1502.05199})

\bibitem{Acciarri:2015uup}
Acciarri R {\em et~al.\/} (DUNE) 2015  (\textit{Preprint} \eprint{1512.06148})

\bibitem{Aartsen:2014oha}
Aartsen M {\em et~al.\/} (IceCube-PINGU Collaboration) 2014  (\textit{Preprint}
  \eprint{1401.2046})

\bibitem{Abe:2011ts}
Abe K {\em et~al.\/} 2011  (\textit{Preprint} \eprint{1109.3262})

\bibitem{Hofestadt:2017jfu}
Hofestädt J (KM3NeT) 2017 {Prospects for measuring the neutrino mass hierarchy
  with KM3NeT/ORCA} {\em {25th European Cosmic Ray Symposium (ECRS 2016) Turin,
  Italy, September 04-09, 2016}\/} (\textit{Preprint} \eprint{1701.04078})
  \urlprefix\url{http://inspirehep.net/record/1509197/files/arXiv:1701.04078.pdf}

\bibitem{An:2015jdp}
An F {\em et~al.\/} (JUNO) 2016 {\em J. Phys.\/} {\bf G43} 030401
  (\textit{Preprint} \eprint{1507.05613})

\bibitem{Mezzacappa:2005ju}
Mezzacappa A 2005 {\em Ann. Rev. Nucl. Part. Sci.\/} {\bf 55} 467--515

\bibitem{Janka:2006fh}
Janka H~T, Langanke K, Marek A, Martinez-Pinedo G and Mueller B 2007 {\em Phys.
  Rept.\/} {\bf 442} 38--74 (\textit{Preprint} \eprint{astro-ph/0612072})

\bibitem{Raffelt:2012kt}
Raffelt G~G 2012 {\em Proc. Int. Sch. Phys. Fermi\/} {\bf 182} 61--143
  (\textit{Preprint} \eprint{1201.1637})

\bibitem{Janka:2012wk}
Janka H~T 2012 {\em Ann. Rev. Nucl. Part. Sci.\/} {\bf 62} 407--451
  (\textit{Preprint} \eprint{1206.2503})

\bibitem{Mirizzi:2015eza}
Mirizzi A, Tamborra I, Janka H~T, Saviano N, Scholberg K, Bollig R, Hudepohl L
  and Chakraborty S 2016 {\em Riv. Nuovo Cim.\/} {\bf 39} 1 (\textit{Preprint}
  \eprint{1508.00785})

\bibitem{Wallace:2015xma}
Wallace J, Burrows A and Dolence J~C 2016 {\em Astrophys. J.\/} {\bf 817} 182
  (\textit{Preprint} \eprint{1510.01338})

\bibitem{Tamborra:2012ac}
Tamborra I, Muller B, Hudepohl L, Janka H~T and Raffelt G 2012 {\em Phys.
  Rev.\/} {\bf D86} 125031 (\textit{Preprint} \eprint{1211.3920})

\bibitem{Huedepohl:2009wh}
Huedepohl L, Muller B, Janka H~T, Marek A and Raffelt G~G 2010 {\em Phys. Rev.
  Lett.\/} {\bf 104} 251101 (\textit{Preprint} \eprint{0912.0260})

\bibitem{sntutorial}
Scholberg K 2018 Neutrinos from supernovae and other astrophysical sources {\em
  The State of the Art of Neutrino Physics: A Tutorial for Graduate Students
  and Young Researchers\/} ed Ereditato A (World Scientific)

\bibitem{Smirnov:2016xzf}
Smirnov A~{\relax Yu} 2016  (\textit{Preprint} \eprint{1609.02386})

\bibitem{Mikheev:1986gs}
Mikheev S~P and Smirnov A~{\relax Yu} 1985 {\em Sov. J. Nucl. Phys.\/} {\bf 42}
  913--917 [Yad. Fiz.42,1441(1985)]

\bibitem{Wolfenstein:1977ue}
Wolfenstein L 1978 {\em Phys. Rev.\/} {\bf D17} 2369--2374

\bibitem{Dighe:1999bi}
Dighe A~S and Smirnov A~Y 2000 {\em Phys. Rev.\/} {\bf D62} 033007
  (\textit{Preprint} \eprint{hep-ph/9907423})

\bibitem{Kuo:1986sk}
Kuo T~K and Pantaleone J~T 1986 {\em Phys. Rev. Lett.\/} {\bf 57} 1805--1808

\bibitem{Fogli:2003dw}
Fogli G~L, Lisi E, Montanino D and Mirizzi A 2003 {\em Phys. Rev.\/} {\bf D68}
  033005 (\textit{Preprint} \eprint{hep-ph/0304056})

\bibitem{Friedland:2006ta}
Friedland A and Gruzinov A 2006  (\textit{Preprint} \eprint{astro-ph/0607244})

\bibitem{Kneller:2010sc}
Kneller J~P and Volpe C 2010 {\em Phys. Rev.\/} {\bf D82} 123004
  (\textit{Preprint} \eprint{1006.0913})

\bibitem{Lund:2013uta}
Lund T and Kneller J~P 2013 {\em Phys. Rev.\/} {\bf D88} 023008
  (\textit{Preprint} \eprint{1304.6372})

\bibitem{Kneller:2013ska}
Kneller J~P and Mauney A~W 2013 {\em Phys. Rev.\/} {\bf D88} 025004
  (\textit{Preprint} \eprint{1302.3825})

\bibitem{Dighe:2003vm}
Dighe A~S, Kachelriess M, Raffelt G~G and Tomas R 2004 {\em JCAP\/} {\bf 0401}
  004 (\textit{Preprint} \eprint{hep-ph/0311172})

\bibitem{Lunardini:2001pb}
Lunardini C and Smirnov A~Y 2001 {\em Nucl. Phys.\/} {\bf B616} 307--348
  (\textit{Preprint} \eprint{hep-ph/0106149})

\bibitem{Fogli:2001pm}
Fogli G~L, Lisi E, Montanino D and Palazzo A 2002 {\em Phys. Rev.\/} {\bf D65}
  073008 [Erratum: Phys. Rev.D66,039901(2002)] (\textit{Preprint}
  \eprint{hep-ph/0111199})

\bibitem{Duan:2010bg}
Duan H, Fuller G~M and Qian Y~Z 2010 {\em Ann. Rev. Nucl. Part. Sci.\/} {\bf
  60} 569--594 (\textit{Preprint} \eprint{1001.2799})

\bibitem{Hannestad:2006nj}
Hannestad S, Raffelt G~G, Sigl G and Wong Y~Y~Y 2006 {\em Phys. Rev.\/} {\bf
  D74} 105010 [Erratum: Phys. Rev.D76,029901(2007)] (\textit{Preprint}
  \eprint{astro-ph/0608695})

\bibitem{Duan:2007bt}
Duan H, Fuller G~M, Carlson J and Qian Y~Z 2007 {\em Phys. Rev. Lett.\/} {\bf
  99} 241802 (\textit{Preprint} \eprint{0707.0290})

\bibitem{Raffelt:2007yz}
Raffelt G~G and Sigl G 2007 {\em Phys. Rev.\/} {\bf D75} 083002
  (\textit{Preprint} \eprint{hep-ph/0701182})

\bibitem{EstebanPretel:2007ec}
Esteban-Pretel A, Pastor S, Tomas R, Raffelt G~G and Sigl G 2007 {\em Phys.
  Rev.\/} {\bf D76} 125018 (\textit{Preprint} \eprint{0706.2498})

\bibitem{Sawyer:2008zs}
Sawyer R~F 2009 {\em Phys. Rev.\/} {\bf D79} 105003 (\textit{Preprint}
  \eprint{0803.4319})

\bibitem{EstebanPretel:2008ni}
Esteban-Pretel A, Mirizzi A, Pastor S, Tomas R, Raffelt G~G, Serpico P~D and
  Sigl G 2008 {\em Phys. Rev.\/} {\bf D78} 085012 (\textit{Preprint}
  \eprint{0807.0659})

\bibitem{duan}
 2010  {Duan H, private communication}

\bibitem{Adams:2013qkq}
Adams C {\em et~al.\/} (LBNE Collaboration) 2013  (\textit{Preprint}
  \eprint{1307.7335})

\bibitem{Duan:2010bf}
Duan H and Friedland A 2011 {\em Phys. Rev. Lett.\/} {\bf 106} 091101
  (\textit{Preprint} \eprint{1006.2359})

\bibitem{Esmaili:2014gya}
Esmaili A, Peres O~L~G and Serpico P~D 2014 {\em Phys. Rev.\/} {\bf D90} 033013
  (\textit{Preprint} \eprint{1402.1453})

\bibitem{Horowitz:2003cz}
Horowitz C~J, Coakley K~J and McKinsey D~N 2003 {\em Phys. Rev.\/} {\bf D68}
  023005 (\textit{Preprint} \eprint{astro-ph/0302071})

\bibitem{Beacom:2001bm}
Beacom J, Boyd R and Mezzacappa A 2001 {\em Physical Review D\/} {\bf 63}
  073011

\bibitem{Arnaud:2001gt}
Arnaud N, Barsuglia M, Bizouard M~A, Cavalier F, Davier M, Hello P and Pradier
  T 2002 {\em Phys. Rev.\/} {\bf D65} 033010 (\textit{Preprint}
  \eprint{hep-ph/0109027})

\bibitem{Langaeble:2016gs}
Langaeble K, Meroni A and Sannino F 2016 {\em Physical Review D\/} {\bf 94}
  053013

\bibitem{Lu:2014zma}
Lu J~S, Cao J, Li Y~F and Zhou S 2015 {\em JCAP\/} {\bf 1505} 044
  (\textit{Preprint} \eprint{1412.7418})

\bibitem{Rossi-Torres:2015rla}
Rossi-Torres F, Guzzo M~M and Kemp E 2015  (\textit{Preprint}
  \eprint{1501.00456})

\bibitem{Drexlin:2005zt}
Drexlin G (KATRIN) 2005 {\em Nucl. Phys. Proc. Suppl.\/} {\bf 145} 263--267
  [,263(2005)]

\bibitem{Kachelriess:2004ds}
Kachelriess M, Tomas R, Buras R, Janka H~T, Marek A and Rampp M 2005 {\em Phys.
  Rev.\/} {\bf D71} 063003 (\textit{Preprint} \eprint{astro-ph/0412082})

\bibitem{evan}
Ott C~D {\em et~al.\/} 2012  [Nucl. Phys. Proc. Suppl.235-236,381(2013)]

\bibitem{Schirato:2002tg}
Schirato R~C and Fuller G~M 2002  (\textit{Preprint} \eprint{astro-ph/0205390})

\bibitem{Dasgupta:2005wn}
Dasgupta B and Dighe A 2007 {\em Phys. Rev.\/} {\bf D75} 093002
  (\textit{Preprint} \eprint{hep-ph/0510219})

\bibitem{Choubey:2006aq}
Choubey S, Harries N~P and Ross G~G 2006 {\em Phys. Rev.\/} {\bf D74} 053010
  (\textit{Preprint} \eprint{hep-ph/0605255})

\bibitem{Kneller:2007kg}
Kneller J~P, McLaughlin G~C and Brockman J 2008 {\em Phys. Rev.\/} {\bf D77}
  045023 (\textit{Preprint} \eprint{0705.3835})

\bibitem{Gava:2009pj}
Gava J, Kneller J, Volpe C and McLaughlin G~C 2009 {\em Phys. Rev. Lett.\/}
  {\bf 103} 071101 (\textit{Preprint} \eprint{0902.0317})

\bibitem{Raffelt:2007cb}
Raffelt G~G and Smirnov A~Y 2007 {\em Phys. Rev.\/} {\bf D76} 081301
  (\textit{Preprint} \eprint{0705.1830})

\bibitem{Raffelt:2007xt}
Raffelt G~G and Smirnov A~Y 2007 {\em Phys. Rev.\/} {\bf D76} 125008
  (\textit{Preprint} \eprint{0709.4641})

\bibitem{Dasgupta:2009mg}
Dasgupta B, Dighe A, Raffelt G~G and Smirnov A~Y 2009 {\em Phys. Rev. Lett.\/}
  {\bf 103} 051105 (\textit{Preprint} \eprint{0904.3542})

\bibitem{Dasgupta:2010cd}
Dasgupta B, Mirizzi A, Tamborra I and Tomas R 2010 {\em Phys. Rev.\/} {\bf D81}
  093008 (\textit{Preprint} \eprint{1002.2943})

\bibitem{Choubey:2010up}
Choubey S, Dasgupta B, Dighe A and Mirizzi A 2010  (\textit{Preprint}
  \eprint{1008.0308})

\bibitem{snowglobes}
  \url{http://www.phy.duke.edu/$\sim$schol/snowglobes}

\bibitem{Dighe:2003jg}
Dighe A~S, Keil M~T and Raffelt G~G 2003 {\em JCAP\/} {\bf 0306} 006
  (\textit{Preprint} \eprint{hep-ph/0304150})

\bibitem{Mirizzi:2006xx}
Mirizzi A, Raffelt G~G and Serpico P~D 2006 {\em JCAP\/} {\bf 0605} 012
  (\textit{Preprint} \eprint{astro-ph/0604300})

\bibitem{Borriello:2012zc}
Borriello E, Chakraborty S, Mirizzi A, Serpico P~D and Tamborra I 2012 {\em
  Phys. Rev.\/} {\bf D86} 083004 (\textit{Preprint} \eprint{1207.5049})

\bibitem{Liao:2016uis}
Liao W 2016 {\em Phys. Rev.\/} {\bf D94} 113016 (\textit{Preprint}
  \eprint{1607.03334})

\bibitem{Dighe:2003be}
Dighe A~S, Keil M~T and Raffelt G~G 2003 {\em JCAP\/} {\bf 0306} 005
  (\textit{Preprint} \eprint{hep-ph/0303210})

\bibitem{Totani:1997vj}
Totani T, Sato K, Dalhed H~E and Wilson J~R 1998 {\em Astrophys. J.\/} {\bf
  496} 216--225 (\textit{Preprint} \eprint{astro-ph/9710203})

\bibitem{Odrzywolek:2003vn}
Odrzywolek A, Misiaszek M and Kutschera M 2004 {\em Astropart. Phys.\/} {\bf
  21} 303--313 (\textit{Preprint} \eprint{astro-ph/0311012})

\bibitem{Wright:2016xma}
Wright W~P, Nagaraj G, Kneller J~P, Scholberg K and Seitenzahl I~R 2016 {\em
  Phys. Rev.\/} {\bf D94} 025026 (\textit{Preprint} \eprint{1605.01408})

\bibitem{Wright:2016gar}
Wright W~P, Kneller J~P, Ohlmann S~T, Roepke F~K, Scholberg K and Seitenzahl
  I~R 2017 {\em Phys. Rev.\/} {\bf D95} 043006 (\textit{Preprint}
  \eprint{1609.07403})

\bibitem{Wright:2017zyq}
Wright W~P, Gilmer M~S, {Fr\"ohlich} C and Kneller J~P 2017  (\textit{Preprint}
  \eprint{1706.08410})

\end{thebibliography}

\ack
The author's research is supported by the Department of Energy Office of Science and the National Science Foundation.  The author thanks the members of the Deep Underground Neutrino Experiment Supernova Burst/Low-Energy Working Group for feedback, and especially Jost Migenda for a careful reading.

\end{document}